\documentclass[ALICE,manyauthors]{cernphprep} 
\usepackage{cernunits,cernchemsym,heppennames2}
\usepackage{cite,fancyvrb}
\usepackage{subfig}
\newif\ifrevtex\revtexfalse
\RequirePackage{lineno}
\providecommand*\eg{e.g.\xspace}
\providecommand*\ie{i.e.\xspace}

\providecommand*\gevc{GeV/$c$\xspace}

\providecommand{\degree}{$^{\rm o}$\xspace}
\providecommand{\s}{$\sqrt{s}$\xspace}
\providecommand{\pt}{\ensuremath{p_{\rm t}}\xspace}
\providecommand{\mt}{\ensuremath{m_{\rm t}}\xspace}
\providecommand{\dedx}{d$E$/d$x$\xspace}
\providecommand{\dndy}{d$N$/d$y$\xspace}

\providecommand{\pp}{pp\xspace}

\providecommand{\sigmaTPC}{$\sigma_{\rm{TPC}-\rm{d}\mathit{E}/\rm{d}\mathit{x}}$\xspace}
\providecommand{\sigmaTOF}{$\sigma_{\rm{TOF-PID}}$\xspace}

\begin{document}
\begin{titlepage}
\PHnumber{2012-131}
\PHdate{10 December 2012}
\title{Measurement of electrons from semileptonic heavy-flavour hadron decays
       in \pp collisions at $\sqrt{s}=7$ TeV}
\Collaboration{ALICE Collaboration%
         \thanks{See Appendix~\ref{authorlist} for the list of collaboration
                      members}}
\ShortAuthor{ALICE Collaboration}
\ShortTitle{Electrons from heavy-flavour decays in \pp collisions at
            \mbox{\s$ = 7$~TeV}}
\begin{abstract}
The differential production cross section of electrons from semileptonic 
heavy-flavour hadron decays has been measured at mid-rapidity 
\mbox{($|y| < 0.5$)} in proton-proton collisions at \mbox{\s$ = 7$~TeV} 
with ALICE at the LHC. Electrons were measured in the transverse 
momentum range \mbox{0.5~$<$ \pt $<$~8~GeV/$c$}. Predictions from a fixed 
order perturbative QCD calculation with next-to-leading-log resummation 
agree with the data within the theoretical and experimental uncertainties.
\end{abstract}
\end{titlepage}
\setcounter{page}{2}

\section{Introduction}
\label{sec:intro}

The measurement of heavy-flavour (charm and beauty) production 
serves as an important testing ground of quantum chromodynamics (QCD), 
the theory of the strong interaction. Because of the
large quark masses, heavy-flavour production in proton-proton 
(\pp) collisions proceeds mainly through initial hard parton-parton 
collisions. Therefore, the production cross sections of charm and
beauty quarks should provide a precision test of perturbative QCD (pQCD)
for all values of transverse momenta \pt. In previous experiments 
with $\rm{p}\bar{\rm{p}}$ collisions at the Tevatron (\s$ = 1.96$~TeV), 
charm production cross sections were measured at high \pt only and
were found to exceed, by about 50\%~\cite{Acosta:2003ax}, the cross
sections expected from pQCD 
calculations~\cite{fonll,fonll2,PhysRevLett.96.012001}. This, however, 
is still compatible with the substantial theoretical uncertainties. 
Beauty production at the Tevatron is well described by such 
calculations~\cite{Acosta:2004yw}.

While the measurement of heavy-flavour production in \pp collisions
is important in its own interest, it also provides a crucial
baseline for corresponding measurements in ultrarelativistic heavy-ion 
collisions. In such collisions a strongly interacting partonic medium 
is formed~\cite{Arsene:2004fa,Adcox:2004mh,Back:2004je,Adams:2005dq}. 
Heavy quarks interact with this medium after they have been 
produced in the initial stage of the collision. Consequently, heavy
quarks suffer energy loss while they propagate through the medium, and 
they participate in the collective dynamics. 
The resulting modifications of the heavy-flavour momentum distributions 
in heavy-ion collisions with respect to those in \pp collisions present 
a sensitive probe for the medium properties~\cite{Adare:2006nq}.

Heavy-flavour production can be investigated, among other channels, via the 
measurement of the contribution of semileptonic heavy-flavour decays to the 
inclusive lepton spectra. Both charm and beauty hadrons have substantial 
branching ratios (\mbox{$\sim 10\%$}) to single electrons or single 
muons~\cite{Nakamura:2010zzi}, giving rise to a large ratio of signal 
leptons from heavy-flavour hadron decays to background from other lepton 
sources, in particular at high \pt. 

Single electrons from heavy-flavour decays were first observed in the
range $1.6 < \pt < 4.7$~\gevc in \pp collisions at the CERN ISR at 
\mbox{\s$ = 52.7$~GeV}~\cite{Busser:1974ej}, before the actual discovery 
of charm.  At the CERN S$\rm{p}\bar{\rm{p}}$S, the UA1 experiment measured 
beauty production via single muons ($10 < \pt < 40$~\gevc) at 
\mbox{\s$ = 630$~GeV}~\cite{Albajar:1990zu} while the UA2 experiment
used single electrons ($0.5 < \pt < 2$~\gevc) to measure the charm 
production cross section~\cite{Botner:1989jg}. 
At the Tevatron, both the CDF and D0 experiments measured beauty 
production via single electrons 
($7 < \pt < 60$~\gevc)~\cite{PhysRevLett.71.500} and single 
muons ($3.5 < \pt < 60$~\gevc)~\cite{Abachi:1994kj}, respectively.

At RHIC, semileptonic heavy-flavour decays were extensively studied
in \pp and, for the first time, in heavy-ion collisions, mainly in the 
electron channel. With the PHENIX experiment the range 
$0.3 < \pt < 9$~\gevc was covered~\cite{Adare:2010de}, and with the STAR 
experiment electrons from heavy-flavour hadron decays were measured in 
the range $3 < \pt < 10$~\gevc~\cite{Agakishiev:2011mr}. 
Within experimental and theoretical uncertainties pQCD calculations are 
in agreement with the measured production cross sections of electrons from 
charm~\cite{Adare:2006hc,Agakishiev:2011mr} and 
beauty decays~\cite{Adare:2009ic,Aggarwal:2010xp} at mid-rapidity in 
\pp collisions at \mbox{\s$ = 0.2$~TeV}. 
In Au-Au collisions, the total yield of electrons from heavy-flavour decays 
was observed to scale with the number of binary nucleon-nucleon 
collisions~\cite{Adler:2004ta}. However, a strong suppression of the electron
yield was discovered for 
\mbox{\pt$ > 2$~GeV/$c$}~\cite{Adler:2005xv,PhysRevLett.106.159902} with a 
simultaneous observation of a nonzero electron elliptic flow strength v$_2$ 
for \mbox{\pt$ < 2$~GeV/$c$}~\cite{Adare:2006nq}, indicating the substantial 
interaction of heavy quarks with the medium produced in Au-Au collisions 
at RHIC.

At the LHC, heavy-flavour production is studied in \pp collisions at higher 
energies. Perturbative QCD calculations agree well with lepton production 
cross sections from heavy-flavour hadron decays measured for 
\mbox{\pt$ > 4$~GeV/$c$} with the ATLAS experiment at 
\mbox{\s$ = 7$~TeV}~\cite{Aad:2011rr}. 
Furthermore, pQCD calculations of beauty hadron decays are in good agreement
with production cross sections of non-prompt J$/\psi$ at mid-rapidity as 
measured with the CMS experiment at high \pt (\mbox{\pt$ > 6.5$~GeV/$c$})
\cite{Khachatryan:2010yr} and with ALICE (A Large Ion Collider Experiment) at
lower \pt (\mbox{\pt$ > 1.3$~GeV/$c$}) \cite{alice_jpsi_b}. 
D-meson production cross sections measured with ALICE are reproduced by 
corresponding calculations within substantial uncertainties at 
7~TeV~\cite{ALICE-D2H} and at 2.76~TeV~\cite{alice_d2h_276}. 
In addition, pQCD calculations are in agreement with the spectra of muons 
from heavy-flavour hadron decays at moderate \pt as measured with 
ALICE at 7~TeV~\cite{muons_in_pp} and at 2.76~TeV~\cite{muons_in_pp_276}. 
It is of particular importance to investigate charm production at 
low \pt~\cite{ALICE-D2H} in order to measure the total charm production cross 
section with good precision. Furthermore, low-\pt charm measurements at the 
LHC probe the parton distribution function of the proton in the region of 
parton fractional momenta $x \sim 10^{-4}$ and squared momentum transfers 
$Q^2 \sim ({\rm 4~GeV})^2$, where gluon saturation effects might play a 
role~\cite{Alekhin:2005dy}.

This paper presents a measurement of single electrons, \mbox{(e$^+$+e$^-$)/2}, 
from semileptonic decays of charm and beauty hadrons in the transverse 
momentum range \mbox{0.5~$<$ \pt $<$~8~GeV/$c$} at mid-rapidity 
\mbox{($|y| < 0.5$)} in \pp collisions at \mbox{\s$ = 7$~TeV} with ALICE. 
For such a measurement an excellent electron identification (eID) and precise 
knowledge of the remaining hadron background in the electron candidate sample 
are mandatory. Two complementary eID approaches are employed. Both are based on
the particle specific energy loss \dedx in the ALICE Time Projection Chamber, 
required to be compatible with the energy loss of electrons. 
To increase the purity of the electron candidate sample, in the first approach
a combination of time-of-flight measurements and the response of the transition
radiation detector is employed (TPC-TOF/TPC-TRD-TOF analysis). In the second 
approach, electromagnetic calorimetry is used (TPC-EMCal analysis).

This article is organised as follows: Section~\ref{sec:alice}
gives an overview over the ALICE detector systems that are relevant
for the analysis presented here. The details of the data analysis are
described in Section~\ref{sec:analysis}. The differential production 
cross section of electrons from semileptonic heavy-flavour decays is 
presented in Section~\ref{sec:results}. In the same Section, pQCD calculations 
at fixed order with next-to-leading-log resummation (FONLL~\cite{fonll,fonll2,fonll3})
are compared with the data, which extend the ATLAS measurement of electrons
from heavy-flavour hadron decays to lower \pt. This article concludes with a
summary in Section~\ref{sec:summary}.

\section{ALICE setup}
\label{sec:alice}

ALICE~\cite{ALICE} is the experiment at the LHC dedicated to the study of 
heavy-ion collisions. The standard ALICE coordinate system is used, in which 
the interaction point (IP) where the particles collide is at the origin of a 
right-handed Cartesian co-ordinate system. From the IP the $z$ axis is along 
the beam pipe, the $x$ axis points towards the center of the LHC, $\phi$ is 
the azimuthal angle around the $z$ axis, and $\theta$ is the polar angle with 
respect to this axis. The setup includes a muon spectrometer at backward 
pseudorapidity \mbox{($-4 < \eta < -2.5$)} and a central barrel comprising 
several detector subsystems located inside a large solenoidal magnet. The 
magnet provides a uniform magnetic field of 0.5~T along the beam direction.
Most of the barrel detectors have a common pseudorapidity coverage of 
\mbox{$-0.9 < \eta < 0.9$}.
The apparatus is described in detail elsewhere~\cite{ALICE}. 
In the following, the detectors used in the analysis are discussed briefly.
For guidance, Fig.~\ref{fig:alice} shows a schematic beam view at $z = 0$ 
of the ALICE central barrel detectors as present during the 2010 running 
period of the LHC.

\ifrevtex
\input{figALICE_revtex.tex}
\else
\begin{figure}[tbh]
\begin{center}
\includegraphics[width=0.7\linewidth]{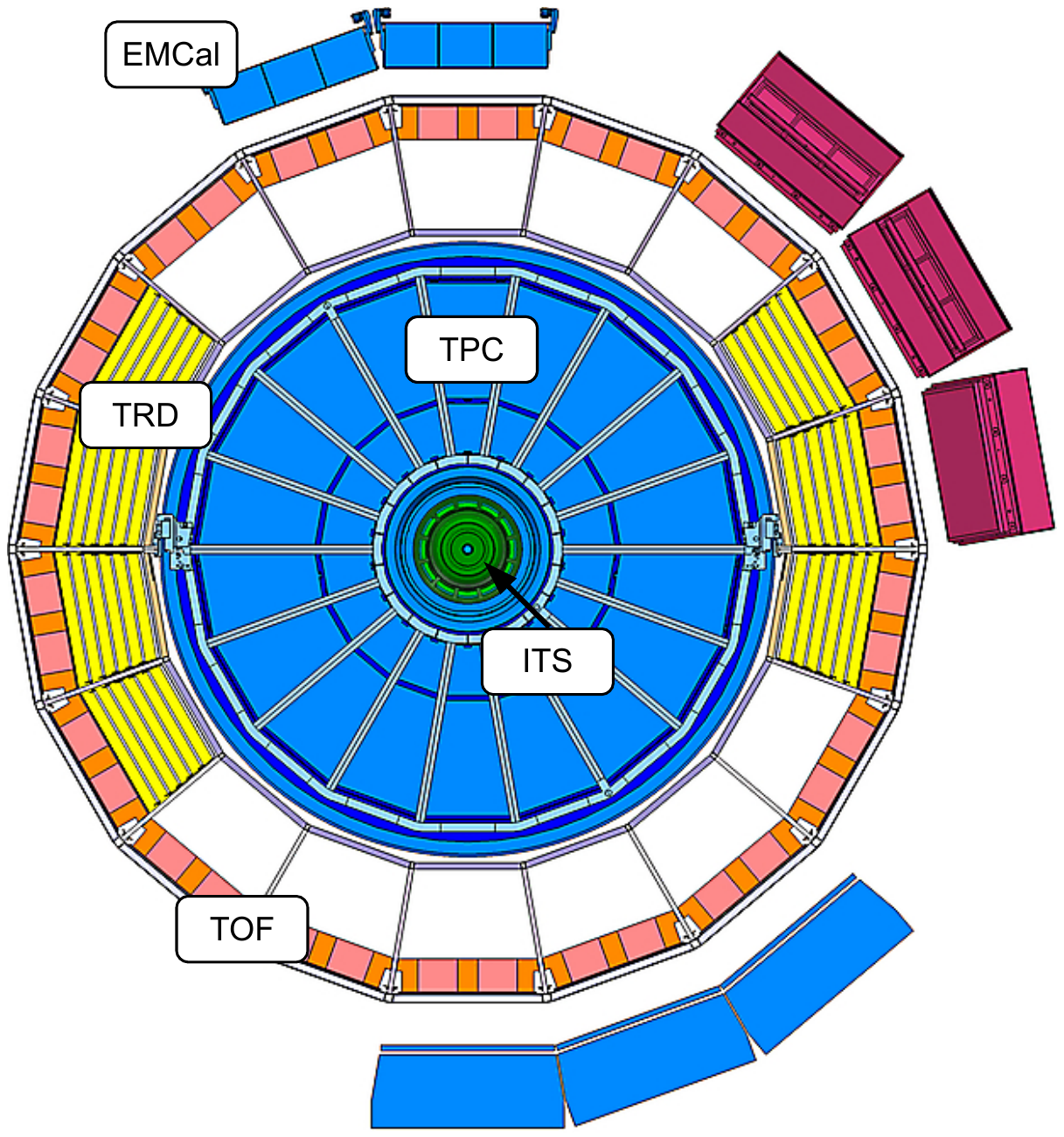}
\end{center}
\caption{(Colour online) Schematic beam view at $z = 0$ of the ALICE central
barrel detectors during the 2010 running period of the LHC. The detectors used
in the present analysis are the Inner Tracking System (ITS), the Time
Projection Chamber (TPC), the Transition Radiation Detector (TRD), the
Time-Of-Flight Detector (TOF), and the Electromagnetic Calorimeter (EMCal).}
\label{fig:alice}
\end{figure}

\fi

The vacuum beam pipe is made of beryllium with a thickness of 800~$\mu$m, and 
an inner diameter  of 58~mm. For protection the pipe is wrapped with polyimide
with a thickness of about 80~$\mu$m. The corresponding material budget is 
0.26\% of a radiation length ($X_{0}$) at $\eta = 0$.

The beam pipe is surrounded by the Inner Tracking System (ITS). The ITS 
provides high-resolution space points for charged particle tracks close to the 
interaction point, thus improving the momentum and angular resolution. 
The ITS includes six cylindrical layers employing three different 
silicon detector technologies. The two innermost layers (at radii of 3.9~cm 
and 7.6~cm), which are equipped with Silicon Pixel Detectors (SPD), provide 
a spatial resolution of 12~$\mu$m in the plane perpendicular to the beam 
direction ($r\phi$) and 100~$\mu$m along the beam axis ($z$). 
About 83\% of the SPD channels were operational for charged particle 
detection during the data taking relevant for this analysis. The SPD also 
contributes to the collision trigger providing a fast estimation of the event 
multiplicity. The two intermediate layers of the ITS are built with 
Silicon Drift Detectors (SDD) and the two outermost layers consist of 
double-sided Silicon Strip Detectors (SSD). Their radii extend from 
15 to 43~cm. The ITS modules were aligned using survey information, 
cosmic-ray tracks, and pp data with the methods described 
in~\cite{Aamodt:2010aa}. The material budget of the entire ITS corresponds 
on average to about 7.18\% of $X_{0}$ at $\eta = 0$~\cite{ALICE}. The exact 
knowledge of the material 
budget in the innermost ITS layers is crucial here as the conversion of 
photons into electron-positron pairs in material is the source of an important 
background component in the present analysis. In the ALICE experiment, the 
reconstruction of such conversion pairs has resulted in a measurement of the 
relevant material budget with a precision of 4.5\%~\cite{Koch:2011fw}.

The most important detector for the track reconstruction and the momentum
measurement is the Time Projection Chamber (TPC), which is also used for 
particle identification~\cite{TPCref}. The ALICE TPC is a large cylindrical 
drift detector whose active volume extends radially from 85 to 247~cm, and 
from -250 to +250~cm along the beam direction. The active volume of nearly 
90~m$^3$ is filled with a Ne (85.5\%), CO$_2$ (9.5\%), and N$_2$ (4.8\%) gas 
mixture. A central high-voltage electrode maintained at -100~kV divides the 
TPC into two sections. The end-caps are equipped with multiwire proportional 
chambers with cathode pad readout. For a particle traversing the TPC, up to 
159 space points (clusters) are recorded. The cluster data are used to 
reconstruct the charged particle trajectory in the magnetic field as 
well as to calculate the particle's specific energy loss \dedx in the TPC gas. 
Simultaneous measurements of the \dedx and momentum allow the identification 
of the particle species which has produced the track. The \dedx resolution of 
the TPC, \sigmaTPC, was approximately 5.5\% for minimum ionising particles 
crossing the full detector~\cite{thesis_kalweit}. The \dedx resolution was 
determined using minimum ionising pions and cosmic ray muons at the Fermi 
plateau. Charged particle tracks are reconstructed in the ITS and 
TPC with a transverse momentum resolution ranging from about 1\% at 1~\gevc, 
to about 3\% at 10~\gevc~\cite{TPCref}.

The TPC is surrounded by the Transition Radiation Detector (TRD) at a radial 
distance of \mbox{2.9~m} from the beam axis. The TRD is segmented in the 
azimuth direction in 18 individual super-modules, seven of which were installed
in the 2010 running period of ALICE as indicated in Fig.~\ref{fig:alice}. 
Each super-module is segmented further in five units (stacks) along the beam 
direction. Each stack comprises six layers in the radial direction.
Each detector element consists of a fibre sandwich radiator of 48~mm thickness
\cite{Radiator}, a drift section of 30~mm thickness, and a multiwire
proportional chamber section (7~mm thickness) with pad readout. The gas is a 
mixture of \mbox{Xe (85\%)} and \mbox{CO$_2$ (15\%)}
\cite{TRDref,PositionReconstruction,PulseHeight,SpaceCharge}.
The scope of the TRD is to provide a good separation of electrons from pions,
particularly for momenta above 1 \gevc. This is accomplished by measuring
transition radiation photons, which are produced only by electrons 
\cite{NeuralNetwork}. The TRD is also designed to provide a fast trigger with 
particle identification information to discriminate electrons from 
hadrons~\cite{TRDTrigger}. This trigger was not used in the 2010 data taking.

At larger radii, at a distance of \mbox{3.7~m} from the beam axis, the 
Time-Of-Flight (TOF) detector provides further essential information for the 
particle identification. The TOF detector is segmented in 18 sectors and covers
the full azimuth. Each sector contains 91 Multigap Resistive Plate Chambers 
(MRPCs). In total, 152,928 sensitive pads of dimension 
\mbox{2.5$\times$3.5 cm$^2$} are read out. The TOF resolution of the particle 
arrival time is, at present, better than 100~ps~\cite{TOF_COMMISSIONING}. 
The start time of the collision is measured by the ALICE T0 detector,
an array of Cherenkov counters located at +350 cm and -70 cm along the
beam-line, or it is estimated using the particle arrival times at the TOF 
detector in events without a T0 signal. In the case that neither of the two 
methods provides an output an average start time is used. 
Depending on the start time method used, the corresponding
resolution is taken into account in the overall TOF PID resolution.
The particle identification is based on the difference between the measured 
time-of-flight and its expected value, computed for each mass hypothesis from 
the track momentum and length of the trajectory. The overall resolution of 
this difference \sigmaTOF is about 160~ps~\cite{ALICE-D2H}.

The Electromagnetic Calorimeter (EMCal) is a Pb-scintillator sampling 
calorimeter, located at a radial distance of 
about \mbox{4.5~m} from the beam line. The full detector covers the 
pseudorapidity range \mbox{$-0.7 < \eta < 0.7$} with an azimuthal acceptance 
of \mbox{$\Delta\phi = 107$\degree}. In the 2010 running period of ALICE the azimuthal 
coverage of the EMCal was limited to \mbox{$\Delta\phi = 40$\degree},
since only part of the detector was installed.
The calorimeter is of the 'Shashlik' type built from alternating lead and 
scintillator segments of 1.44~mm and 1.76~mm thickness, respectively, together
with longitudinal wavelength-shifting fibres for light collection. The cell 
size of the EMCal is approximately 0.014~$\times$~0.014~rad in 
$\Delta \phi \times \Delta \eta$, and the depth corresponds to 20.1 X$_0$. 
From electron test beam data, the energy resolution of the EMCal was
determined to be
$1.7\bigoplus 11.1/\sqrt{E({\rm GeV})}\bigoplus 5.1/E({\rm GeV})\%$~\cite{EMCref}.  

A minimum \pt of about 0.3~\gevc is needed for the particles to reach the
TRD, TOF, and EMCal detectors in the magnetic field of 0.5~T.

The VZERO detector is used for event selection and background rejection.
It consists of two arrays of 32 scintillators each, which are arranged in  
four rings around the beam pipe on either side of the interaction region,
covering the pseudorapidity ranges \mbox{$2.8 < \eta < 5.1$} and 
\mbox{$-3.7 < \eta < -1.7$}, respectively. 
The time resolution of this detector is better than 1~ns. Information from the 
VZERO response is recorded in a time window of $\pm$~25~ns around the nominal 
beam crossing time. The VZERO is used to select beam-beam interactions in the 
central region of ALICE and to discriminate against interactions of the beam 
with gas molecules in the beam pipe.

The ALICE minimum bias trigger required at least one hit in either of the two 
SPD layers or in the VZERO detector. In addition, collision events had 
to be in coincidence with signals from the beam position monitors,
indicating the passage of proton bunches 
from both beams.

\section{Analysis}
\label{sec:analysis}

\subsection{General strategy}
\label{subsec:stratgey}
For the measurement of the differential invariant cross section of electrons 
from semileptonic decays of heavy-flavour hadrons the following strategy was 
adopted.
First, charged particle tracks which fulfil a set of electron identification 
cuts were selected. From the electron candidate tracks the remaining 
contamination from misidentified hadrons was subtracted. After corrections 
for geometrical acceptance and efficiency the inclusive electron yield per 
minimum bias triggered collision was determined for two different electron 
identification strategies. Since for all relevant sources the spectra 
of decay positrons and electrons are identical (${\rm e}^{+}/{\rm e}^{-} = 1$),
the average spectrum of positrons and electrons, $({\rm e}^{+} + {\rm e}^{-})/2$,
was used for the further analysis. The electron background from sources other 
than semileptonic heavy-flavour hadron decays was calculated using a cocktail 
approach and subtracted from the inclusive electron spectra. 
The resulting spectra of electrons from heavy-flavour hadron decays were 
normalised using the cross section of minimum bias triggered \pp collisions. 
A weighted average of the two measurements obtained with different electron 
identification strategies led to the final result.

\subsection{Data set and event selection}
\label{subsec:events}
The data used in the present analysis were recorded during 
the 2010 running period.
The luminosity was limited to $0.6 - 1.2 \times 10^{29} \rm{cm}^{-2}\rm{s}^{-1}$
in order to keep the probability of collision pile-up per triggered event
below $2.5\%$. This was cross-checked by looking at events with more than 
one vertex reconstructed with the SPD. 

The primary collision vertex can be determined using the reconstructed 
tracks in the event or the correlated hits in the two pixel layers. Only 
events with a reconstructed primary vertex using one of the two methods were 
selected for further analysis. In order to minimise edge effects at the limit of the 
central barrel acceptance, the vertex was required to be within 
\mbox{$\pm 10$~cm} from the centre of the ALICE experiment along the beam 
direction. Integrated luminosities of 2.6~$\rm{nb}^{-1}$ and 2.1~$\rm{nb}^{-1}$ 
were used for the TPC-TOF/TPC-TRD-TOF and TPC-EMCal analysis, respectively.

In the offline analysis, pile-up events were identified using the SPD.
Events with a second interaction vertex reconstructed with at least three 
tracklets (short tracks from SPD clusters)
and well separated from the first vertex by more than 8 mm, 
are rejected from further analysis.
Taking into account the efficiency of the pile-up event identification, 
less than 2.5\% 
of the triggered events have been found to be related to more than one 
interaction. The effect of the remaining undetected pile-up 
was negligible for the analysis.
Moreover, background from beam-gas interactions was eliminated using the VZERO 
timing information as well as the correlation in the SPD between the number of 
reconstructed charged particle track segments and the number of hits.

\subsection{Track reconstruction and selection}
\label{subsec:tracks}
\ifrevtex
\input{tableTracks_revtex}
\else
\begin{table}[t]
\begin{center}
\caption{Track selection cuts: except for the cut on the
number of ITS hits and the request for hits in the SPD,
the selections were common to all analysis strategies.
See text for details.}
\label{tab::trackcuts}
\begin{tabular}{l|l}
\hline\hline
Track property & Requirement \\
\hline
Number of TPC clusters                               & $\geq$ 120 \\
Number of TPC clusters used in the \dedx calculation & $\geq$ 80 \\
Number of ITS hits in TPC-TOF/TPC-TRD-TOF            & $\geq$ 4 \\
Number of ITS hits in TPC-EMCal                      & $\geq$ 3 \\
SPD layer in which a hit is requested in TPC-TOF/TPC-TRD-TOF & first \\ 
SPD layer in which a hit is requested in TPC-EMCal & any \\ 
$\chi^2$/ndf of the momentum fit in the TPC          & $<$ 2 \\
Distance of Closest Approach in $xy$ (cm)             & $<$ 1 \\
Distance of Closest Approach in $z$ (cm)              & $<$ 2 \\
\hline\hline
\end{tabular}
\end{center}
\end{table}

\fi

Charged particle tracks reconstructed in the TPC and ITS were propagated 
towards the outer detectors using a Kalman filter approach~\cite{TrackingKF}. 
Geometrical matching was applied to associate tracks with hits in the outer 
detectors. 

In the currently limited active area in azimuth of the TRD, the tracks were 
associated with track segments, called tracklets, reconstructed in individual 
chambers. This tracklet reconstruction assumed straight trajectories 
of charged particles passing a chamber. As the ALICE TRD comprises six
layers, a track can include up to six tracklets. In the TPC-TRD-TOF 
analysis a minimum of four associated TRD tracklets was required for each 
electron candidate track. For each tracklet the charge deposited in the 
corresponding chamber was measured. This information was used for electron 
identification. 

The EMCal coverage was limited in the 2010 run. In azimuth, the installed EMCal 
sectors neither overlap with the installed TRD supermodules nor with the area 
of the innermost SPD layer which was operational in 2010 as indicated in
Fig.~\ref{fig:alice}.
Electromagnetic showers reconstructed in the EMCal were associated with charged 
particle tracks if the distance between the track projection on the EMCal 
surface and the reconstructed shower was small in $\eta$\xspace and 
$\phi$\xspace. The quadratic sum of the 
difference between track projection and reconstructed position had to be less 
than 0.05 in $(\eta$,$\phi)$ space for a track-shower pair to be accepted, where $\phi$ is measured in radians.

The pseudorapidity ranges used in the TPC-TOF/TPC-TRD-TOF and TPC-EMCal
analyses were restricted to \mbox{$|\eta| < 0.5$} and \mbox{$|\eta| < 0.6$}, 
respectively, because towards larger absolute values of $\eta$ the systematic
uncertainties related to particle identification increase considerably. 


Electron candidate tracks were required to fulfil several track selection cuts. 
Table~\ref{tab::trackcuts} summarises these selection criteria.
A cut on the $\chi^2$ per degree of freedom (ndf) of the momentum fit in the TPC
was applied to reject fake tracks which comprise a significant number of 
clusters originating from more than one charged particle trajectory. 
A track reconstructed within the TPC is characterised by the number of clusters
used for the track reconstruction and fit (up to a maximum of 159 clusters).
Not all of these clusters are used for the energy loss calculation:
those close to the borders of the TPC sectors are not considered.
Separate cuts are applied on these two quantities.
To guarantee good particle identification based on the specific \dedx
in the TPC, tracks were required to include a minimum number of 80 clusters
used for the energy loss calculation. A cut on the number of clusters for
tracking is used to enhance the electron/pion separation.
As the energy deposit of electrons on the Fermi plateau is approximately 
1.6 times larger than for minimum ionizing particles, the associated 
clusters are insensitive to detector threshold effects and electron tracks have,
on average, a higher number of clusters. 
The stringent request for at least 120 clusters from the maximum of 159
enhances electrons relative to hadrons.


Kink candidates, \ie tracks which are not 
consistent with the track model of continuous particle trajectories but
show deviations due to decays in flight or the emission of Bremsstrahlung, 
were discarded from further analysis since the \dedx resolution of the TPC 
is worse for such kink tracks than for regular tracks.
In order to minimise the contribution from photon conversions in the ITS, 
a hit in the innermost SPD layer was required for all selected tracks in the
TPC-TOF/TPC-TRD-TOF analysis. In total, at least four ITS hits were required 
to be associated with a track. Since the active area in azimuth of the EMCal 
overlapped with an inactive area of the first SPD layer, this approach
had to be modified for the TPC-EMCal analysis. For the latter case, a matching 
hit was required in any of the two SPD layers and the required total number 
of ITS hits was reduced to three. Charged pion tracks from the weak decay 
\mbox{K$^0_{\rm S} \rightarrow \pi^+ \pi^-$}
occurring beyond the first SPD layer were used to demonstrate that the 
probability of random matches between tracks and uncorrelated
hits in the ITS is negligible.
A cut on the distance of closest approach (DCA) to the primary vertex in the
transverse plane ($xy$) as well as in the beam direction ($z$) was applied to 
reject background tracks and non-primary tracks.

\subsection{Electron identification}
\label{subsec:eID}
\ifrevtex
\input{figTPCdEdx_revtex}
\else
\begin{figure}[t]
\begin{center}
\includegraphics[width=0.49\linewidth]{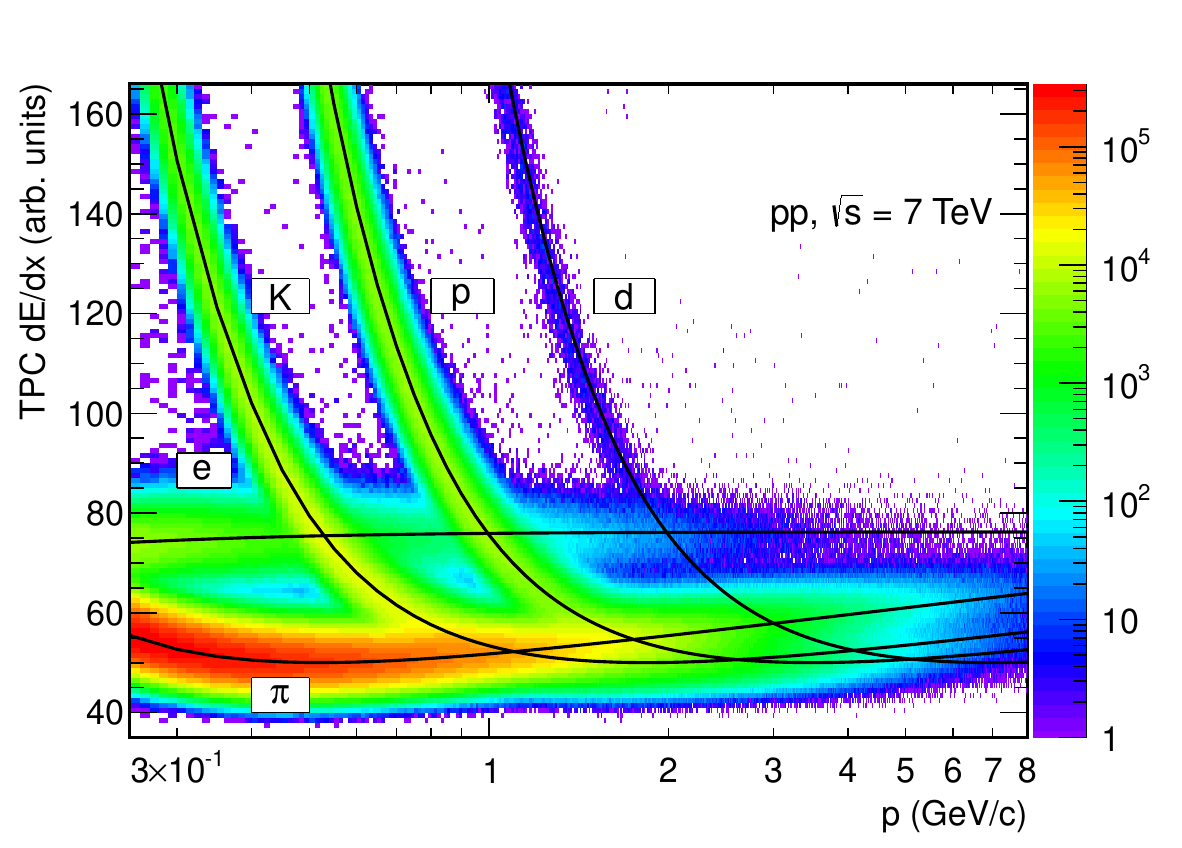}
\includegraphics[width=0.49\linewidth]{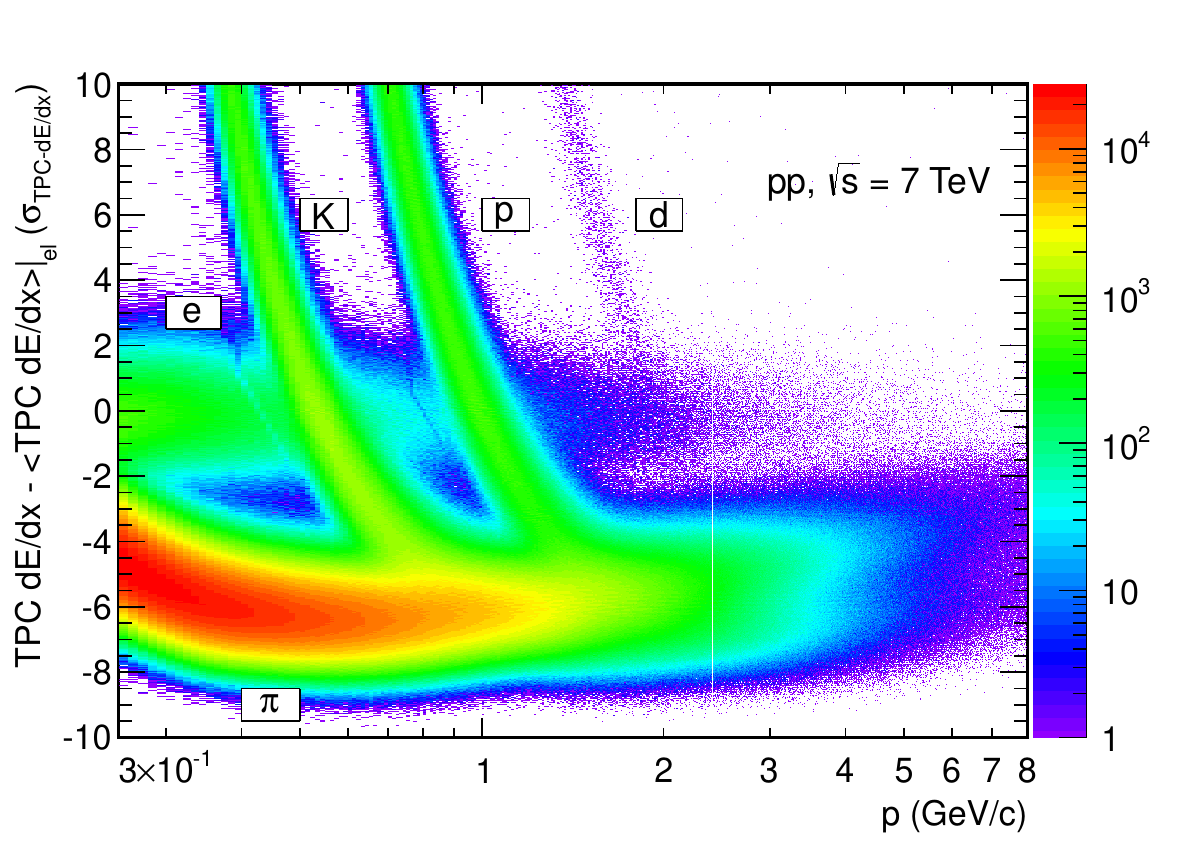}
\end{center}
\caption{(Colour online) Specific energy loss \dedx in arbitrary units
measured in the TPC as a function of the reconstructed charged particle momentum
(left panel), and expressed as a deviation from the expected energy loss of
electrons, normalised by the energy loss resolution (right panel). Contributions
from both positively and negatively charged particles are included.}
\label{dedx}
\end{figure}

\fi

Electrons were identified using the information provided by various detector 
subsystems of the ALICE central barrel. The detector which played
the most important role in particle identification for both analyses discussed 
here is the TPC. Particle identification in the TPC is based on the measurement
of the specific energy loss \dedx in the detector gas.
The \dedx distribution, expressed in arbitrary units, as a function of the 
particle momentum for tracks measured in 7~TeV \pp collisions, is shown in the 
left panel of Fig.~\ref{dedx}. The solid lines depict the energy loss for 
electrons, pions, kaons, protons, and deuterons expected from the Bethe-Bloch 
formula~\cite{BetheBloch}.
For the electron selection, the energy loss was expressed as a deviation from
the parameterised electron Bethe-Bloch line, divided by the energy loss 
resolution \sigmaTPC, as shown in the right panel of Fig.~\ref{dedx}.


Figure~\ref{dedx} demonstrates that the electron identification provided by 
the TPC is not sufficient at low momentum (below 1.5~\gevc) because the kaon
and proton \dedx lines cross the electron line.
In addition, the merging of the \dedx lines of electrons, muons, pions, and 
other hadrons limits the particle identification at high momentum. 
Therefore, a high purity electron candidate sample can only be selected with 
the help of other detectors. Two different strategies were used in this 
analysis, one employing in addition the information from the TOF and TRD 
detectors, and the other one based on the EMCal response. 

\subsubsection{TPC-TOF/TPC-TRD-TOF analysis}
The information provided by the TOF detector is complementary to
that from the TPC in the low momentum region and it
is used to resolve the ambiguities
in the crossing regions of the TPC electron, kaon, and proton lines. The 
time-of-flight information allows the rejection of kaons up to a momentum
of approximately 1.5~\gevc and protons up to about 3~\gevc. 
The selection was done by comparing the measured time-of-flight with 
the value expected assuming the particle being an electron. Only tracks 
compatible with the electron hypothesis within 3~\sigmaTOF were
considered as electron candidates for further analysis.
The difference between the measured time-of-flight and the expected time-of-flight,
as a function of the momentum, is shown in the upper left panel of
Fig.~\ref{fig::elid::PIDcuts}. Lines indicate the selection band.
This criterion combined with the selection of tracks between 0 and 3~\sigmaTPC 
resulted in a pure sample of electron candidates up to a momentum of 
approximately 4~\gevc. In this momentum range, the hadron contamination remained
below 1\%, while
above 4~\gevc the pion contamination became significant 
again. At such high momenta the TOF information could not be used to
reduce further the hadron contamination in the electron candidate sample. Therefore, 
the TPC-TOF analysis was restricted to the \pt range below 4~\gevc.
To extend the accessible range to higher momenta, information from
the TRD was used. As for the TPC, particle identification 
in the TRD makes use of
the specific energy loss in the detector gas. In
addition, the measurement of transition radiation photons produced by 
electrons traversing the dedicated radiators in front of the TRD drift 
chambers enhances distinctively the capability of the TRD to separate electrons 
from hadrons.
The charge deposit per tracklet was compared with reference charge 
distributions obtained from dedicated test beam data~\cite{Bailhache:2006}, 
where electron and pion beams were provided at a number of different, 
discrete momenta. The probability of identifying a particle of given momentum 
as an electron was derived from a linear interpolation between the nearest 
measured data points in momentum. The electron probabilities were calculated 
for each TRD tracklet (up to six per track). They were combined for a given 
track and a likelihood value was calculated on which the eID is based.

\ifrevtex
\input{figPID_revtex.tex}
\else
\begin{figure}[t]
\includegraphics[width=.98\textwidth]{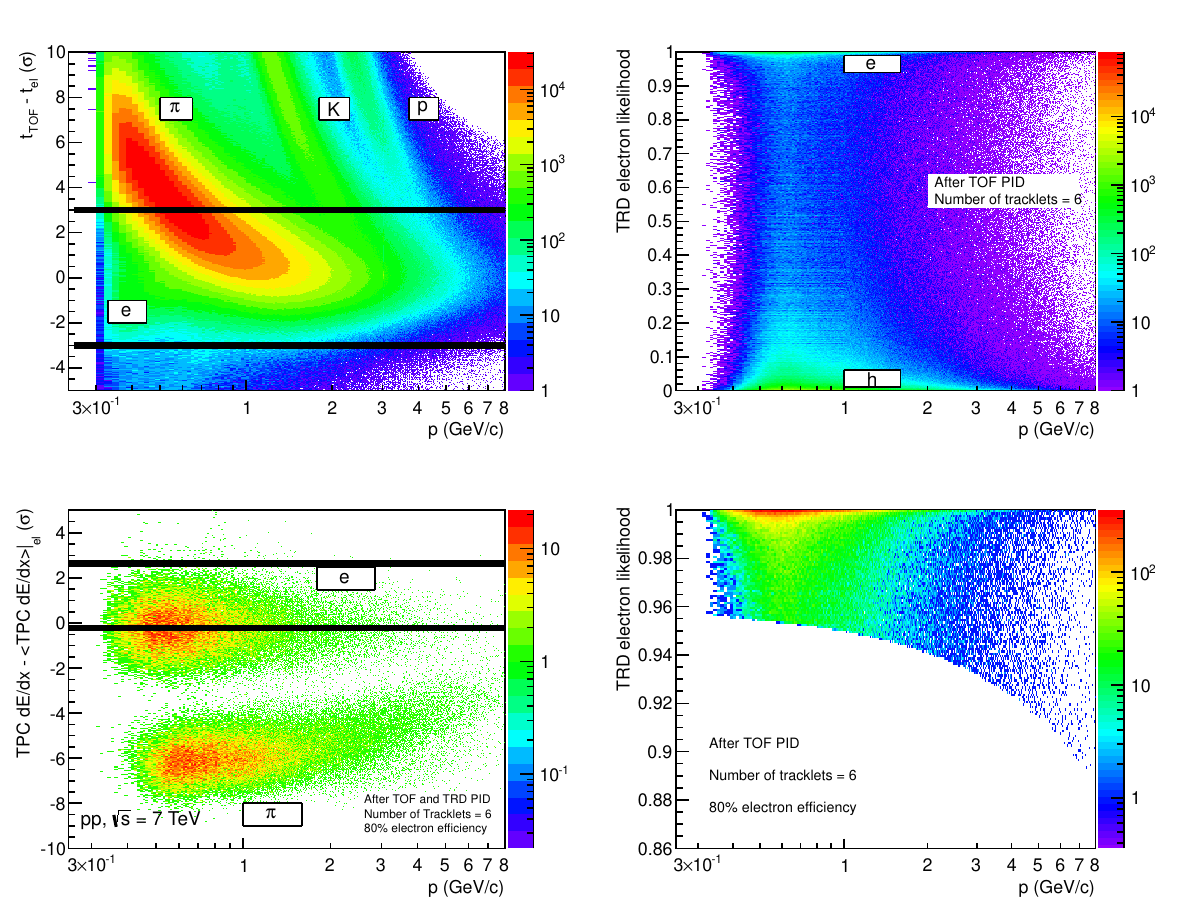}
\caption{(Colour online) Electron selection with the TOF, TRD, and TPC detectors.
The difference between measured and expected time-of-flight is shown in the upper left panel. Lines indicate the selection band. For tracks selected by TOF,
the TRD electron likelihood distribution for tracks with 6 TRD tracklets is
shown in the upper right panel. The lower right panel displays the TRD electron
likelihood distribution for tracks with an electron efficiency of 80\% in the
TRD (note the compressed scale on the vertical axis). For tracks passing the
TRD selection, the TPC \dedx, expressed in units of the \dedx resolution
(\sigmaTPC) is shown in the lower left panel. Lines indicate the electron
selection band. The parameterisation of the expected energy loss of electrons
in this data period, and the specific selection criteria of this analysis are
such that the mean (width) of the electron \dedx distribution is not exactly
zero (one). Therefore, the selection band is slightly shifted from the nominal
values of 0 and 3~\sigmaTPC.}
\label{fig::elid::PIDcuts}
\end{figure}

\fi

The TRD electron likelihood distribution as a function of momentum for tracks 
passing the TOF selection and having six TRD tracklets is shown in the upper 
right panel of Fig.~\ref{fig::elid::PIDcuts}. The electron candidate 
selection was performed applying a momentum dependent cut defined such that it 
provided a constant electron efficiency of 80\%. The \pt dependence of this
cut was determined using a clean sample of electrons from photon conversions.
Furthermore, this cut depends on the exact number of charge measurements 
(tracklets) available per track (four to six in the present analysis). 
The lower right panel of  Fig.~\ref{fig::elid::PIDcuts} depicts the cut 
described for six tracklets. 
Cuts for tracks with four or five tracklets were applied in the same way. 
The TRD selection was applied only for tracks with a momentum above 4~\gevc
because at lower momenta the TPC-TOF selection was sufficient.
For tracks passing the TRD selection, the lower left panel of 
Fig.~\ref{fig::elid::PIDcuts} shows the particle \dedx in the TPC, 
expressed as the distance to the expected energy deposit of electrons, 
normalised by the energy loss resolution. Having used the TRD information,
an excellent separation of electrons from pions is already visible in the whole 
momentum range up to 8~\gevc. The selection of tracks between 0 and 3~\sigmaTPC
results in an almost pure sample of electrons with a remaining hadron 
contamination of less than 2\% over the full \pt range (see below).

\subsubsection{TPC-EMCal analysis}
An alternative approach to separate electrons from hadrons, over a wide 
momentum range, is based on electromagnetic calorimetry. Tracks were 
geometrically matched  with clusters reconstructed in the EMCal. For each 
track, the momentum information was provided by the track reconstruction 
algorithms in the TPC and ITS. The corresponding energy deposit $E$ was 
measured in the EMCal. The energy information was provided by a cluster of 
cells: the energy deposition was summed over adjacent cells, with an energy 
measurement above a threshold of \mbox{$\approx$48~MeV} around a seed cell. 

For the TPC-EMCal analysis, tracks between $-1.5$ and 3 \sigmaTPC were 
selected. For those candidate tracks, the ratio $E/p$ of the energy deposited 
in the EMCal and the measured momentum was calculated to identify electrons.
The distribution of $E/p$ is shown in Fig.~\ref{fig:eop} for tracks with 
transverse momenta in the range \mbox{$4 < \pt < 5$~\gevc}.
Electrons deposit their total energy in the EMCal and, due to their small mass,
the ratio $E/p$ should be equal to unity. Therefore, the peak around one 
in Fig.~\ref{fig:eop} confirms the good pre-selection of electron candidate 
tracks using the TPC. The exact shape of the $E/p$ distribution depends on the 
EMCal response, Bremsstrahlung in the material crossed by electrons along their
trajectory, and the remaining background from charged hadrons. 
The $E/p$ distribution was fitted with the sum of a Gaussian and an exponential 
function. Electron candidates were required to have $E/p$ between $-3$ and 
$+3$ $\sigma_{E/p}$ of the $E/p$ distribution, where $\sigma_{E/p}$ is the
width of the fitted Gaussian function. Due to the loose ITS cuts, the TPC-EMCal 
analysis suffers from a large background from photon conversions and,
consequently, a small signal to background ratio for electrons from 
heavy-flavour hadron decays at low \pt. Therefore, the \pt range
was limited to $\pt > 3$~\gevc, where a significant heavy-flavour signal
could be measured.
\ifrevtex
\input{figPIDemcal_revtex.tex}
\else
\begin{figure}[tbh]
\begin{center}
\includegraphics[width=0.7\linewidth]{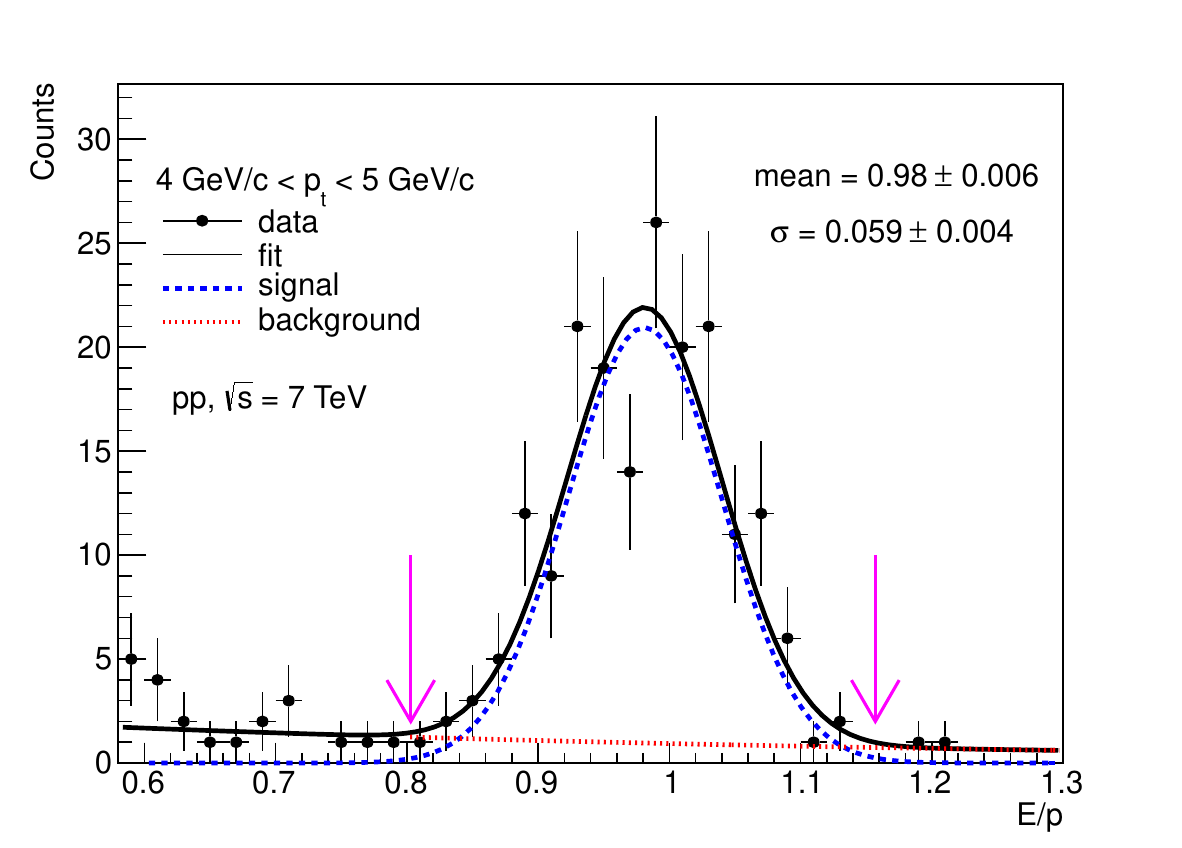}
\end{center}
\caption{(Colour online) Ratio $E/p$ of the energy deposit in the EMCal and the
measured momentum for charged particle tracks in the range
\mbox{$4 < \pt < 5$~\gevc}. The distribution was fitted with the sum of a
Gaussian for the electron signal and an exponential for the remaining hadron
background. Arrows indicate the selection window for electron candidates.}
\label{fig:eop}
\end{figure}

\fi

\subsection{Hadron contamination}
\label{subsec:contamination}
The residual hadron contamination, after the electron identification cuts, was 
estimated by fitting the measured detector signal distributions with functions 
modelling the background and signal contributions. The hadron contamination is summarised in 
Table~\ref{hadroncontamination} for the three analysis strategies.

\ifrevtex
\input{tableeID_revtex.tex}
\else
\begin{table}
\begin{center}
\caption{Overview over the hadron contamination subtracted in the inclusive electron spectrum for the three analysis strategies.}
\label{hadroncontamination}
\begin{tabular}[ht]{l|c|c|c}
\hline\hline
Analysis & TPC-TOF & TPC-TRD-TOF & TPC-EMCal \\
\pt range (\gevc)& 0.5 -- 4 & 4 -- 8 & 3 -- 7 \\
\hline
\hline
Hadron contamination (\%) & $\leq$ 1 & $\leq$ 2 & 7$\pm$4 (sys)\\
\hline\hline
\end{tabular}
\end{center}
\end{table}

\fi

\subsubsection{TPC-TOF/TPC-TRD-TOF analysis}
For the TPC-TOF/TPC-TRD-TOF analysis, the TPC \dedx distribution after TOF- 
and TRD-PID cuts was fitted in momentum slices. 
The residual contamination to the electron sample is given by the contribution of 
misidentified charged particles after the cut on the TPC \dedx. 
The cut on the TPC \dedx applied for electrons was chosen to have 50\% 
efficiency for all momenta. 
The electron line was parameterised using a Gaussian function, which
describes well the shape of the TPC \dedx distribution, expressed as 
deviation from the parameterised electron Bethe-Bloch line 
normalised by the energy loss resolution, for a given 
particle species close to the maximum of this distribution. The dominant 
contribution to the contamination of the electron candidate sample at momenta 
above 1~\gevc comes from the tail of the pion \dedx distribution.
This tail is not adequately described by a Gaussian for the purpose of an 
estimation of the contamination. A better description of the tail of
the pion \dedx distribution is obtained by multiplying a Landau distribution
with an exponential term. The validity of this approach was confirmed using
a clean pion sample from ${\rm K}^{0}_{\rm S}$ decays which was selected using 
the V0-finder and tagged using topological cuts~\cite{V0}. At low momenta, 
protons and kaons are suppressed by the eID cut applied using the TOF detector, 
while at higher momenta the kaon and proton \dedx lines approach each other. 
Therefore, a single slightly skewed Gaussian distribution was used to fit the 
combined contribution of both particle types. The contribution of muons was 
fitted jointly with that of the pions.

\ifrevtex
\input{figContamination_revtex.tex}
\else
\begin{figure}
\centering
\includegraphics[scale=0.7]{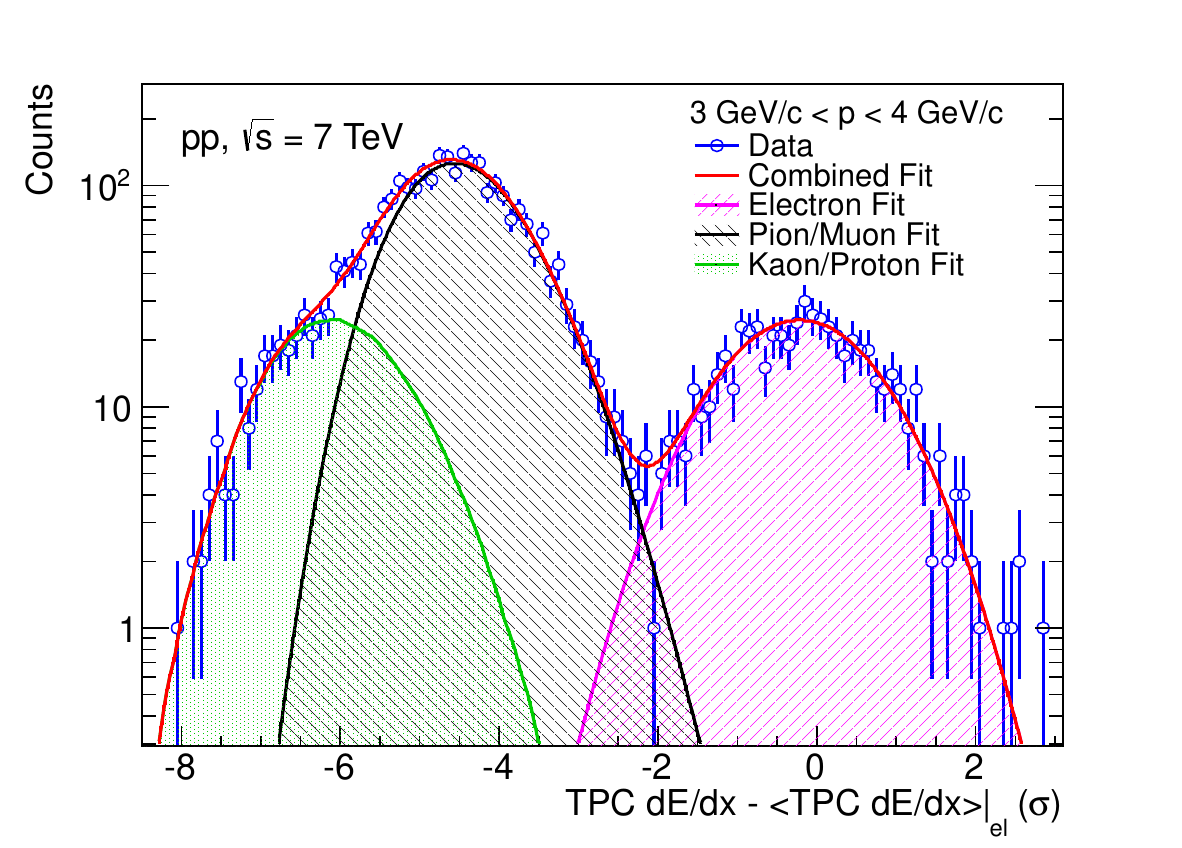}
\caption{(Colour online): The specific energy loss distribution measured with the TPC in the
momentum range \mbox{$3 < p < 4$~\gevc} (histogram) is compared to the sum of
functions describing the contributions from different particle
species. Data and fit agree within statistical uncertainties.}
\label{fig::contfit}
\end{figure}

\fi

The combined fit of the TPC \dedx distribution in the momentum range  
\mbox{$3 < p < 4$~\gevc} is shown in Fig.~\ref{fig::contfit}.
To demonstrate that the fit does not introduce any additional systematic 
uncertainty, the difference between data and fit was compared with the 
expected statistical fluctuations. The fit is in good agreement with the 
data within statistical uncertainties.

The relative contamination was calculated as the ratio of the fitted background 
contribution to the overall distribution after the TPC \dedx cut. 
The contamination remained insignificant (below 2\%) up to a momentum of 
8~\gevc, and it was subtracted from the electron candidate sample in the
TPC-TOF/TPC-TRD-TOF analysis.

\subsubsection{TPC-EMCal analysis}
For the TPC-EMCal analysis, the hadron contamination in the electron candidate 
sample was estimated based on fits to the $E/p$ distribution in momentum slices 
with a function describing the signal (Gaussian for $E/p \sim 1$) and 
background (exponent) as shown in Fig.~\ref{fig:eop}. 
Furthermore, the contamination has been constrained with the ratio of the 
integrals of the $E/p$ distribution in two intervals: 
$\mu_{E/p}$ to \mbox{$\mu_{E/p} + n \cdot \sigma_{E/p}$} and 
\mbox{$\mu_{E/p} - n \cdot \sigma_{E/p}$} to $\mu_{E/p}$ for $n=3$, 
where $\mu,\sigma$ are the parameters of the Gaussian and $\mu_{E/p}$ is the 
mean of the distribution. This ratio is sensitive to the amount of background 
in the measured $E/p$ and its evolution has been studied by varying $n$ 
between 1 and 3. Based on these estimates the hadron contamination in the 
electron candidate sample was determined to be (7$\pm$4)\% in the range 
\mbox{$3 < \pt < 7$~\gevc}, and it was subtracted from the electron sample.

\subsection{Corrections and normalization}
\label{subsec:acc_eff}
Corrections were applied to the electron candidate spectra for the geometrical 
acceptance of the detectors ($\epsilon^{\rm geo}$), the reconstruction efficiency 
($\epsilon^{\rm reco}$), and the electron identification efficiency 
($\epsilon^{\rm eID}$). 

\ifrevtex
\input{figEfficiency_revtex.tex}
\else
\begin{figure}[h]
\begin{center}
\includegraphics[width=0.78\linewidth]{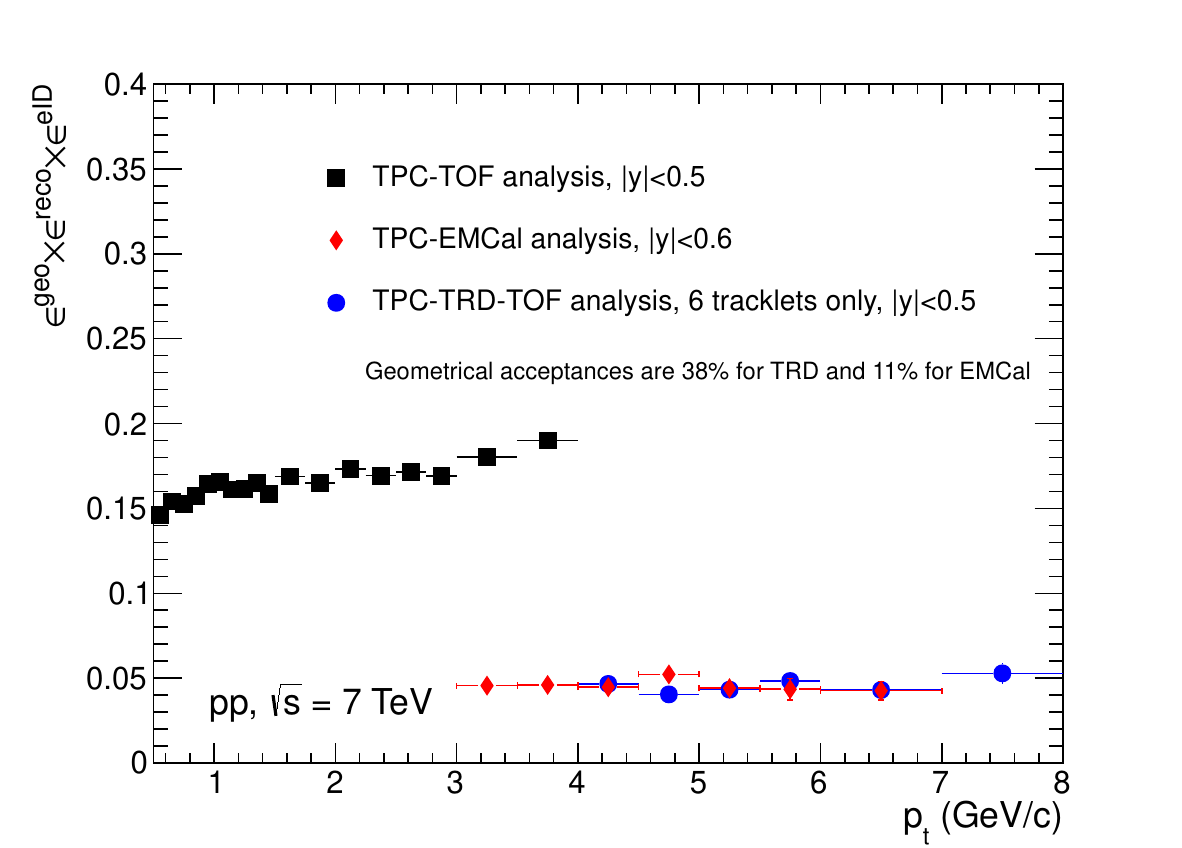}
\end{center}
\caption{(Colour online) Acceptance, tracking, and particle identification
efficiency for electrons at mid-rapidity in \pp collisions at 7~TeV
for the TPC-TOF/TPC-TRD-TOF and the TPC-EMCal analysis. For transverse
momenta below 4~\gevc the TRD was not used for eID. The total
reconstruction efficiency for electrons with the TPC-TRD-TOF eID
approach is shown for the requirement of 6 tracklets in the TRD
as an example.
\label{fig:efficiency}}
\end{figure}

\fi
\ifrevtex
\input{figEfficiencyV0_revtex.tex}
\else
\begin{figure}[h]
\center
\includegraphics[width=0.8\linewidth]{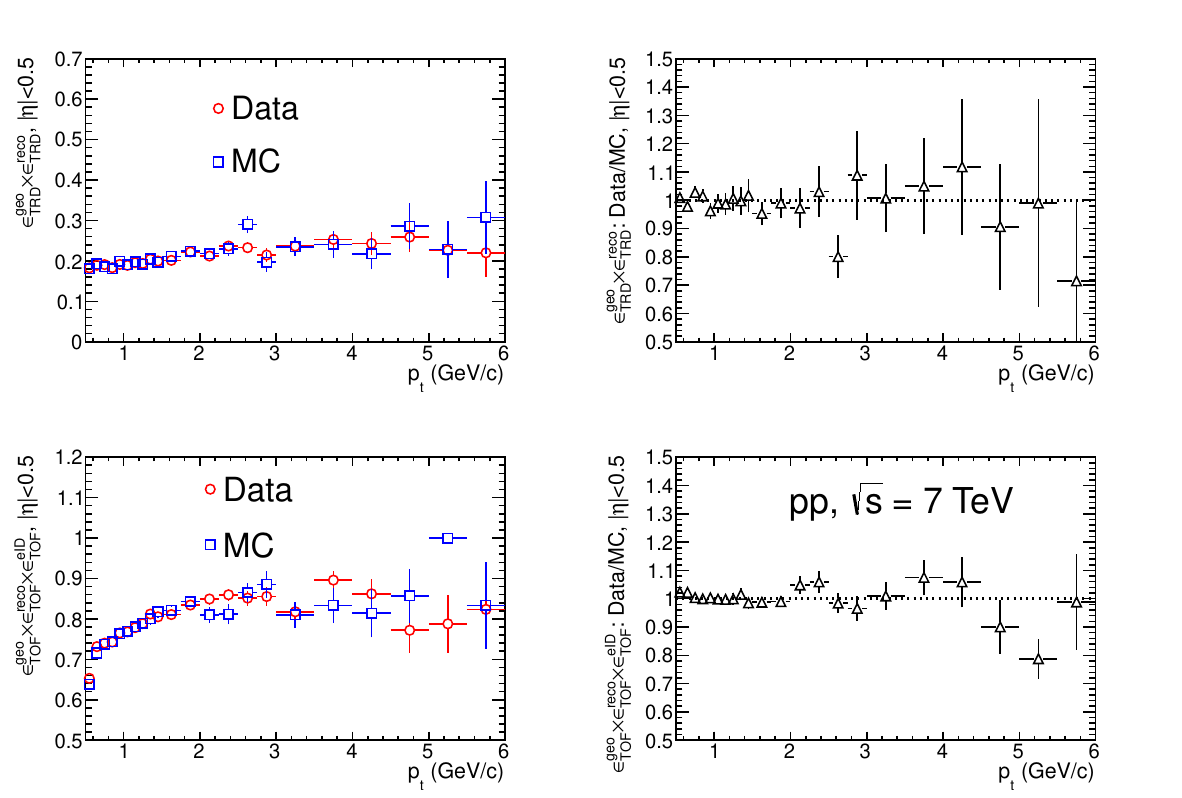}
\caption{(Colour online) Acceptance, tracking, and particle identification
efficiencies are compared in data and in simulation for electrons from photon
conversions in material. Upper panel: TRD acceptance times tracking efficiency
(at least five reconstructed tracklets were required for this example). For
transverse momenta below 4~\gevc the TRD was not used for eID. Lower panel:
TOF matching efficiency times particle identification efficiency.
\label{fig:TPCTOFTRD_efficiency_comparisonMCdata}}
\end{figure}

\fi

%

Due to the finite azimuthal angle covered by the TRD and the EMCal detectors 
in the 2010 run, the maximum geometrical acceptance was 38\% for the 
TPC-TRD-TOF analysis and 11\% for the TPC-EMCal analysis. The geometrical 
acceptance and reconstruction efficiency were computed from a full numerical 
Monte Carlo simulation of the experiment. Monte Carlo events were produced by 
the PYTHIA 6.4.21 event generator~\cite{Sjostrand:2006za} using the Perugia-0 
parameter tuning~\cite{Skands:2009zm} with the same primary vertex distribution
as in the data. The generated particles were propagated through the apparatus 
using GEANT3~\cite{Geant3}. The same reconstruction algorithms and cuts were 
used as for the analysis of data.
For the calculation of $\epsilon^{\rm geo}$ and $\epsilon^{\rm reco}$ in the
TPC-TOF/TPC-TRD-TOF analysis, which requires a hit in the first SPD layer,
only those electrons were considered in the simulation which were produced
within 3~cm distance from the interaction vertex in the transverse direction 
and which were reconstructed in the pseudorapidity range $|\eta| < 0.5$.
For the TPC-EMCal analysis, which requires a hit in any of the two SPD layers,
electrons produced within 7~cm transverse distance from the vertex and
with $|\eta| < 0.6$ were considered for the calculation of $\epsilon^{\rm geo}$ 
and $\epsilon^{\rm reco}$.

The evaluation of the electron transverse momentum is affected by the finite 
momentum resolution and by the electron energy loss due to Bremsstrahlung in 
the material in front of and in the tracking detectors, where the dominant 
contribution is from the ITS ($X/X_{0} = 7.18$\% at $\eta = 0$~\cite{ALICE}).
These effects distort the shape of the \pt distribution, which falls steeply
with increasing momentum, and have to be taken into account.
The necessary correction grows with increasing steepness
of the \pt distribution and with increasing widths of the \pt bins. 
To determine this correction, an unfolding procedure based on Bayes' 
theorem was applied. The Monte Carlo generated and reconstructed 
transverse momentum distributions of electrons were used to obtain a smearing 
matrix. A detailed description of the procedure can be found 
elsewhere~\cite{Unfolding}. The maximum unfolding correction of the measured
electron yield was $\approx 20$\% at $\pt = 2$~\gevc, becoming smaller towards
higher \pt.

The product of the overall acceptance and efficiency ($\epsilon^{\rm geo}$ $\times$ $\epsilon^{\rm reco}$ 
$\times$ $\epsilon^{\rm eID}$) as function of \pt for the TPC-TOF/TPC-TRD-TOF 
analysis as well as the overall efficiency for the TPC-EMCal analysis are shown in Figure~\ref{fig:efficiency}. 

To cross check the value of the acceptance times efficiency calculated via the simulation 
and to determine TRD PID efficiencies, a data-driven method was employed. 
A pure sample of electrons from photon conversions in the detector material 
was selected. 
Reconstructed conversion electron vertices were selected using the
V0-finder~\cite{V0}. The same cuts as in the analysis were applied 
to the pure electron sample except for the requirements in the ITS which were 
relaxed such that the electron candidates needed to have only two hits in the 
ITS, from which at least one is required to be in any of the two pixel layers.
The cross-check was done in the momentum range where the sample of electrons
from photon conversions is statistically significant (up to 6~\gevc).
The good agreement of the TRD acceptance and tracking efficiency 
($\epsilon^{\rm geo}_{\rm TRD}$ $\times$ $\epsilon^{\rm reco}_{\rm TRD}$) for electrons 
from conversions in data and in the simulation, which have at least five TRD 
tracklets, is demonstrated in 
Fig.~\ref{fig:TPCTOFTRD_efficiency_comparisonMCdata}. 
The TOF tracking and PID efficiency after the TRD requirement 
($\epsilon^{\rm geo}_{\rm TOF}$ $\times$ $\epsilon^{\rm reco}_{\rm TOF}$
$\times$ $\epsilon^{\rm eID}_{\rm TOF}$) 
is also well reproduced in the simulations (see 
Fig.~\ref{fig:TPCTOFTRD_efficiency_comparisonMCdata}). 

For the TPC-EMCal analysis, the electron identification efficiency from the 
TPC \dedx cut was estimated using the data driven method. Particles were
selected with a \dedx in the range between $-1.5$ and 3 $\sigma_{\rm TPC-dE/dx}$. 
The corresponding efficiency was about 93\% with respect to the full 
distribution. 
The efficiency of the electron identification with EMCal, \ie track matching 
and eID employing the $E/p$ cut, was estimated using the simulation.

The \pt-differential invariant yield $N^{\rm{e}^{\pm}}$ of inclusive electrons, 
$(\rm{e}^+ + \rm{e}^-)/2$, was calculated from the corrected electron \pt
spectrum and the number $N_{\rm MB}$ of minimum bias \pp collisions as:
 
\begin{equation}
\frac{1}{2\pi\pt} \frac{{\rm{d}}^{2}N^{\rm{e}^{\pm}}}{{\rm d}\pt{\rm d}y} = 
\frac{1}{2} 
\frac{1}{2\pi\pt^{\rm{centre}}} 
\frac{1}{\Delta y \Delta \pt}
\frac{N_{\rm raw}^{\rm{e}^{\pm}}(\pt)}{(\epsilon^{\rm geo} \times \epsilon^{\rm reco} \times \epsilon^{\rm eID})}
\frac{1}{N_{\rm MB}} \; ,
\label{eq:cross_section}
\end{equation}

where $\pt^{\rm{centre}}$ are the centres of the \pt bins with widths $\Delta \pt$
chosen here, and $\Delta y$ is the width of the rapidity interval covered. In 
the following, invariant yields or cross sections within a given \pt bin are 
always quoted at the bin centre without a bin-shift correction for the fact 
that the electron yield decreases with increasing \pt. When ratios of yields
or cross sections are calculated the same \pt bins are used for both the
numerator and the denominator and average yields or cross sections are used
for every individual \pt bin to avoid bin-shift effects.

\subsection{Systematic uncertainties}
\label{subsec:sys}

\subsubsection{TPC-TOF-TRD analysis}
\label{subsubsec:sys_tpctoftrd}
The following sources of systematic uncertainties were considered: the corrections of the ITS, 
TPC, TOF, and TRD tracking efficiencies, the TOF, TPC, and TRD particle 
identification efficiencies, the \pt unfolding procedure, and the 
absolute normalisation. 

\ifrevtex
\input{tableCutVariation_revtex.tex}
\else
\begin{table}
\begin{center}
\caption{Variation of the electron selection criteria to estimate the 
systematic uncertainties due to track reconstruction and particle 
identification.}
\label{sys_cutvariation}
\begin{tabular}[h]{l|c|c|c}
\hline\hline
Variable & Looser criteria & Reference criteria & Stronger criteria \\
\hline
& & & \\
{\bf{All analyses:}}& & & \\
N. of TPC tracking clusters & $\ge$ 100  & $\ge$ 120 & $\ge$ 140 \\
N. of TPC PID clusters & $\ge$ 80 & $\ge$ 80 & $\ge$ 100, $\ge$ 120 \\
DCA to the primary vertex & $<$ 2 cm ($<$ 4 cm) & $<$ 1 cm ($<$ 2 cm) & $<$ 0.5 cm ($<$ 1 cm) \\
in $xy$ ($z$) &  &  & $<$ 0.3 cm ($<$ 0.5 cm) \\
 &  &  &  \\
{\bf{TPC-TOF and}}  &  &  &  \\
{\bf{TPC-TRD-TOF analyses:}} &  &  &  \\
Number of ITS hits & $\ge$ 3 & $\ge$ 4 & $\ge$ 5\\
TOF compatibility with & $\le$ 4 $\sigma_{\rm TOF-PID}$ & $\le$ 3 $\sigma_{\rm TOF-PID}$  & $\le$ 2 $\sigma_{\rm TOF-PID}$  \\
e hypothesis &  &  &  \\
TPC \dedx cut & -0.254 $<\sigma_{\rm TPC-dE/dx}<$ 3 & 0 $<\sigma_{\rm TPC-dE/dx}<$ 3 & 0.126 $<\sigma_{\rm TPC-dE/dx}<$ 3 \\
              & -0.126 $<\sigma_{\rm TPC-dE/dx}<$ 3 &  & 0.254 $<\sigma_{\rm TPC-dE/dx}<$ 3 \\
 &  &  &  \\
{\bf{TPC-TRD-TOF analysis:}} &  &  &  \\
Fixed electron efficiency  & 85\% & 80\% & 75\% \\
 for TRD likelihood cut &  &  &  \\
 &  &  &  \\
{\bf{TPC-EMCal analysis:}} &  &  &  \\
Number of ITS hits & $\ge$ 2 & $\ge$ 3 & $\ge$ 4\\
TPC \dedx cut & -2 $<\sigma_{\rm TPC-dE/dx}<$ 3 & -1.5 $<\sigma_{\rm TPC-dE/dx}<$ 3 & -1.5 $<\sigma_{\rm TPC-dE/dx}<$ 2 \\
$E/p$ matching & $|\sigma_{E/p}|<$ 4 & $|\sigma_{E/p}|<$ 3 & $|\sigma_{E/p}|<$ 2\\
& & & \\
\hline\hline
\end{tabular}
\end{center}
\end{table}
\fi

To estimate the contributions from tracking and particle identification,
the analysis was repeated with modified selection criteria as summarised 
in Table~\ref{sys_cutvariation}.

For each variation of the selection criteria, the inclusive electron spectra 
were fully corrected. The resulting spectra were compared by inspecting 
their ratio. As function of \pt, these ratios define the relative systematic 
uncertainties as listed in Table~\ref{sys_tpctoftrdoverview}. 
A general systematic uncertainty of 2\%,
due to the ITS-TPC track matching efficiency, was 
taken from dedicated tracking investigations. It is important to note that for 
each cut related to the particle identification the hadron contamination may 
change and has to be re-evaluated.

In addition, the corrected spectra of positrons and electrons, as well as the 
corrected spectra obtained in the positive ($\eta^+$) and negative $\eta$ 
($\eta^-$) range, were compared. Deviations from the expected ratios 
${\rm e}^{+}/{\rm e}^{-} = 1$ and $\eta^{+}/\eta^{-} = 1$ were taken into 
account in the systematics.

The systematic uncertainty related to the MC \pt-distribution used for the 
corrections, named "unfolding" in Table~\ref{sys_tpctoftrdoverview}, was 
extracted from the comparison of the data corrected with two different 
Monte Carlo samples. In addition to the PYTHIA 6.4.21 based sample, used
already for the evaluation of the geometrical acceptance and the reconstruction 
efficiency (see Sec.~\ref{subsec:acc_eff}), a second PYTHIA based sample with
artificially enhanced heavy-flavour hadron yields was employed.

Up to electron transverse momenta of 4~\gevc the electron identification was
based on the TPC-TOF selection only. For higher momenta the TRD selection 
was included. Therefore, the TRD contribution to the systematic uncertainties 
was only considered for the part of the spectrum above 4~\gevc.

The systematic uncertainties are summarised in 
Table~\ref{sys_tpctoftrdoverview}. The systematic uncertainty of the DCA cuts increases at 
low \pt, where the DCA resolution decreases and electrons from photon 
conversion in the material do not point to the primary vertex. The total 
systematic uncertainty is calculated as the quadratic sum of all contributions 
and it is of the order of 8.5\% for the TPC-TOF and between 20\% and 26\% for 
the TPC-TRD-TOF parts of the spectrum, respectively. 

\ifrevtex
\input{tableSysError_revtex.tex}
\else
\begin{table}
\begin{center}
\caption{Overview over the contributions to the systematic uncertainties 
on the inclusive electron spectrum for the three analysis strategies.}
\label{sys_tpctoftrdoverview}
\begin{tabular}[ht]{l|c|c|c}
\hline\hline
Analysis & TPC-TOF & TPC-TRD-TOF & TPC-EMCal \\
\pt range (\gevc)& 0.5 -- 4 & 4 -- 8 & 3 -- 7 \\
\hline
\hline
Error source & \multicolumn{3}{c}{systematic uncertainty [\%]}\\
\hline
Track matching & $\pm$2 & $\pm$2 & $\pm$2\\
ITS number of hits & \pt $<$ 1.0\gevc: +4,-2 & $\pm$5 & $\pm$10\\
                   & \pt $>$ 1.0\gevc: $\pm$2 &  & \\
TPC number of tracking clusters & \pt $<$ 1.1\gevc: +3,-6 & \pt $<$ 6\gevc: $\pm$5 & $\pm$4\\
                                & \pt $>$ 1.1\gevc: $\pm$3 & \pt $>$ 6\gevc: $\pm$4 & \\
TPC number of PID clusters & $\pm$2 & $<$ $\pm$1 & $\pm$2 \\ 
DCA to the primary vertex in $xy$ ($z$) & \pt $<$ 0.6\gevc: +0.5,-2 & $<$ $\pm$1 & \\
                                    & \pt $>$ 0.6\gevc: +0.5 & $<$ $\pm$1 & $<$  $\pm$1\\
TOF matching and PID & $\pm$5 & $\pm$5 & -- \\
TPC PID & $\pm$3 & 4\gevc $<$ \pt $<$ 8\gevc: $\pm$10 & $\pm$6\\
		&		 & \pt $>$ 8\gevc: $\pm$16.7 & \\ 
TRD tracking and PID & -- & $\pm$5 & -- \\
EMCal PID & -- & -- & $\pm$5 \\ 
Electric charge & $\pm$2 & $\pm$ 10 & $\pm$ 10\\
$\eta$ & $\pm$2 & $\pm$ 10 & $\pm$ 10\\
Unfolding & $\pm$3 & $\pm$5 & $\pm$5\\
\hline\hline
\end{tabular}
\end{center}
\end{table}

\fi

\subsubsection{TPC-EMCal analysis}
\label{subsubsec:sys_emcal}
Systematic uncertainties from the electron identification on the inclusive 
electron spectrum obtained with the TPC-EMCal approach arise from the \dedx 
measured in the TPC and the $E/p$ matching.
The uncertainties were estimated by measuring the spectra with changing cuts 
on \dedx and $E/p$. 
The variation of the cuts are summarised in Table~\ref{sys_cutvariation}.
The resulting uncertainty of the electron identification is 5\% from the
$E/p$ matching, which includes the subtraction of contamination, and 6\% from 
the \dedx selection.
The systematic uncertainties due to the track selection were estimated by 
applying the same variation of cuts as for the TPC-TOF/TPC-TRD-TOF analysis, 
except for the ITS cut.
The individual contributions are summarised in Table~\ref{sys_tpctoftrdoverview}.
The total systematic uncertainty is approximately 20\% on the inclusive 
electron spectrum.

\subsection{Inclusive invariant \pt-differential electron yield}
\label{subsec:inclusive}
The electron yield per minimum bias \pp collision was measured as function of 
\pt. The hadron contamination was subtracted statistically from 
the spectrum and corrections for acceptance, reconstruction, and electron 
identification efficiency were applied. The corrected inclusive electron 
spectra measured with the TPC-TOF and TPC-TRD-TOF analyses are shown in 
the upper panel of Fig.~\ref{fig:ince}. The spectra were parameterised
simultaneously using a Tsallis function as depicted in Fig.~\ref{fig:ince}.
The results from both analyses agree well in the \pt region between 1 and 
4~\gevc as demonstrated in the lower panel of Fig.~\ref{fig:ince}, which
shows the ratios of the data to the common fit on a linear scale.
However, the systematic uncertainties in the TPC-TOF analysis are substantially
smaller than in the TPC-TRD-TOF analysis.
Therefore, for the combined TPC-TOF/TPC-TRD-TOF inclusive yield the TPC-TOF 
result is used for $\pt < 4$~\gevc. The extension towards higher \pt is
given by the TPC-TRD-TOF measurement. 
The corresponding result employing the TPC-EMCal eID is also depicted in 
Fig.~\ref{fig:ince}. 
Since the relevant material budget was not the same for the two approaches the 
contribution from photon conversions is different and, hence, the inclusive 
electron yield is larger for the TPC-EMCal analysis than for the 
TPC-TOF/TPC-TRD-TOF analysis.

\ifrevtex
\input{figInclusive_revtex.tex}
\else
\begin{figure}[htbp]
\centering
\includegraphics[width=.55\textwidth]{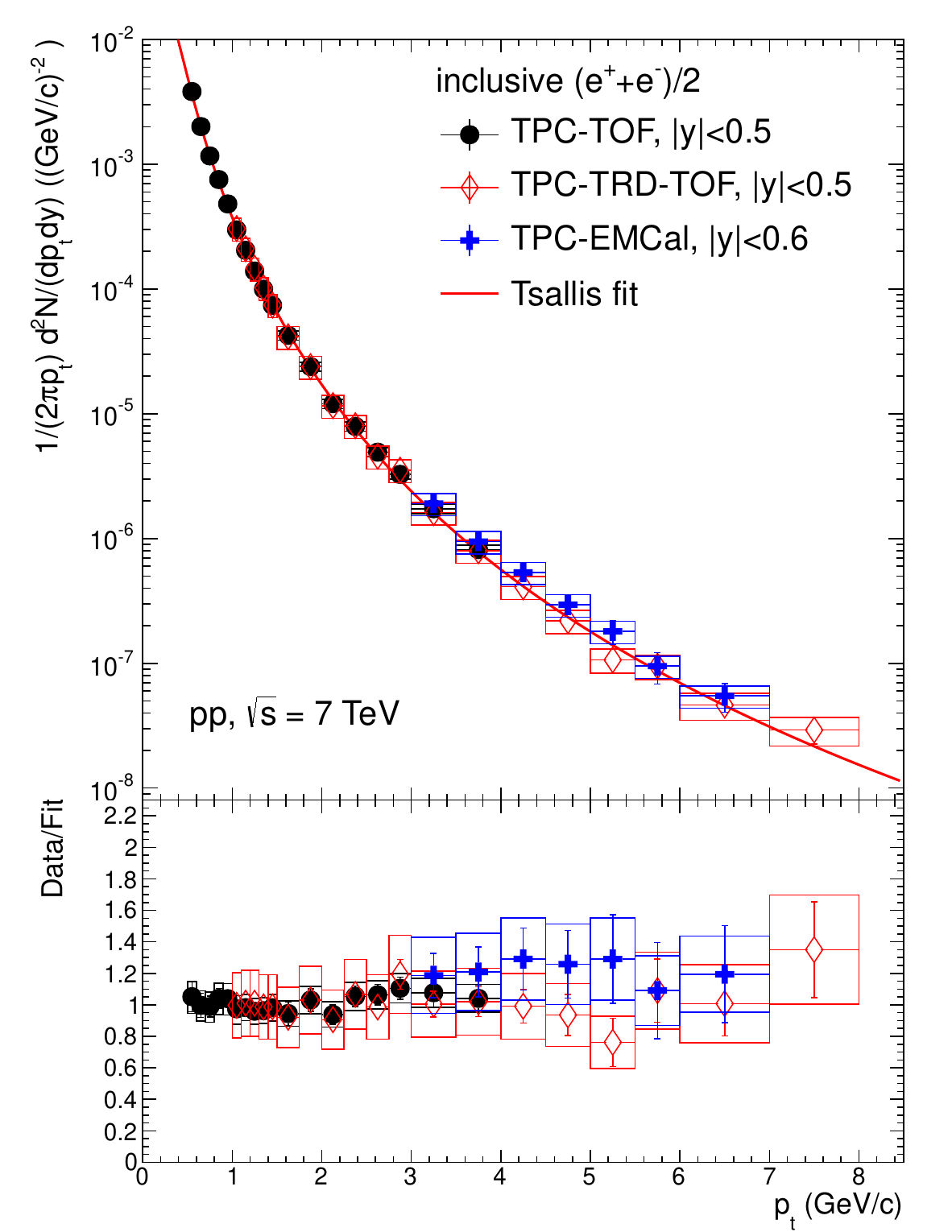}
\caption{(Colour online) Inclusive electron yield per minimum bias collision
as function of \pt measured at mid-rapidity showing the TPC-TOF, TPC-TRD-TOF,
and TPC-EMCal results, respectively, in \pp collisions at
\mbox{\s$ = 7$~TeV}. The TPC-TOF and TPC-TRD-TOF spectra have been
parameterised simultaneously using a Tsallis function (upper panel).
The ratio of the measured spectra to the Tsallis fit is shown in the lower
panel. Statistical uncertainties are indicated by error bars, while
systematic uncertainties are shown as boxes.}
\label{fig:ince}
\end{figure}

\fi

\subsection{Electron background cocktail}
\label{subsec:cocktail}
The inclusive electron spectrum can be subdivided into five components:

\begin{enumerate}
\item signal heavy-flavour electrons, \ie electrons from semileptonic decays 
of hadrons carrying a charm or beauty quark or antiquark,
\item background electrons from Dalitz decays of light neutral mesons and 
from the conversion of decay photons in the material in the detector acceptance,
\item background electrons from weak \mbox{K $\rightarrow$ e$\pi\nu$} 
(K$_{\rm e3}$) decays 
and dielectron decays of light vector mesons,
\item background electrons from dielectron decays of heavy quarkonia 
(J/$\psi$, $\Upsilon$),
\item background electrons originating from partonic hard scattering processes.
This includes electrons from the Drell-Yan process and electrons related to the 
production of prompt photons, \ie both virtual prompt photons 
(electron-positron pairs) as well as real prompt photons which can convert in 
the material of the detector.
\end{enumerate}

Of the background contributions listed above, the first one (Dalitz electrons
and photon conversions in material) is the largest in electron yield. Towards 
high electron \pt, contributions from hard scattering processes (prompt 
photons, decays of heavy-quarkonia, and Drell-Yan processes) are important 
and will, eventually, become dominant.

The signal of electrons from heavy-flavour decays is small compared to the
background at low \pt but rises with increasing \pt as will be shown in
Section~\ref{sec:results} (Fig.~\ref{fig:incl_vs_cocktail}). One technique to 
extract the heavy-flavour signal from the inclusive electron spectrum is the 
so-called ``cocktail subtraction" method described in detail here. In this 
approach, a cocktail of electrons from different background sources was 
calculated using a Monte Carlo hadron-decay generator which, by construction, 
produces identical spectra for decay positrons and electrons. The resulting 
background spectra were then subtracted from the inclusive electron spectrum. 
This approach relies on the availability of the momentum distributions of the 
relevant background sources.

\ifrevtex
\input{figCocktail_revtex.tex}
\else
\begin{figure}
\begin{center}
\includegraphics[width=0.48\linewidth]{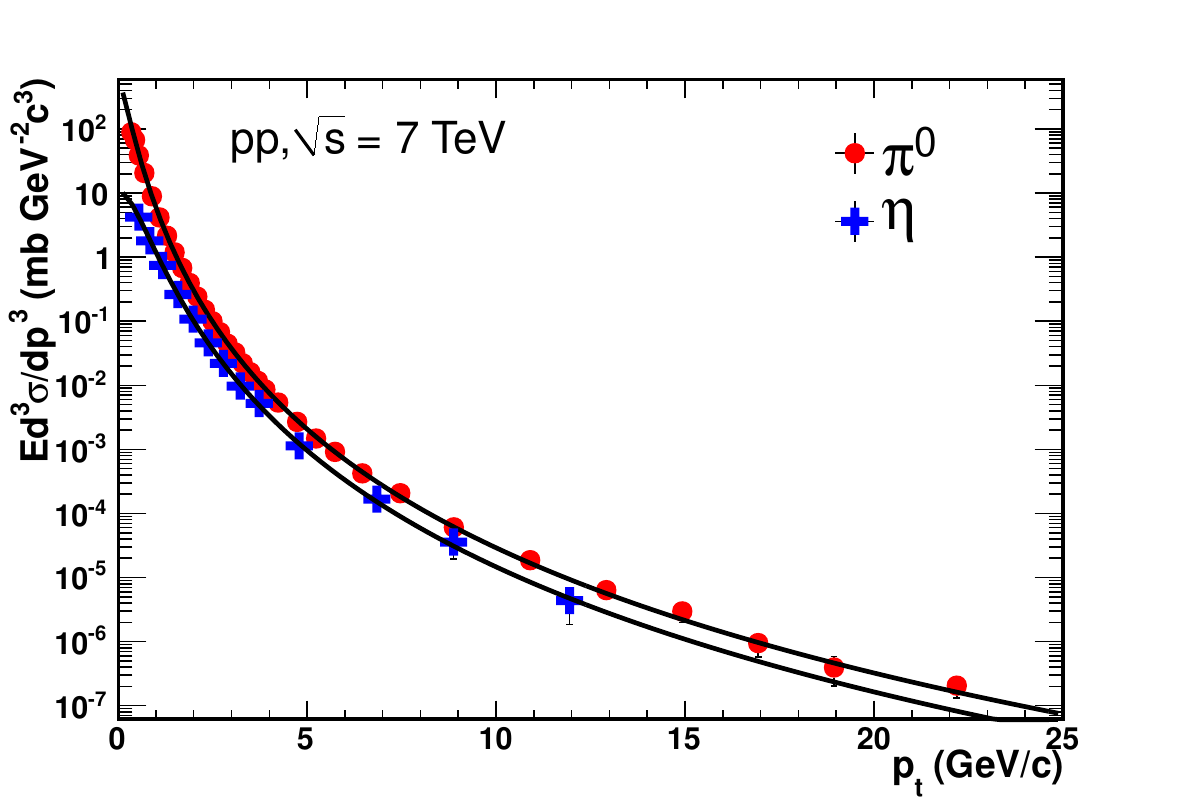}
\includegraphics[width=0.48\linewidth]{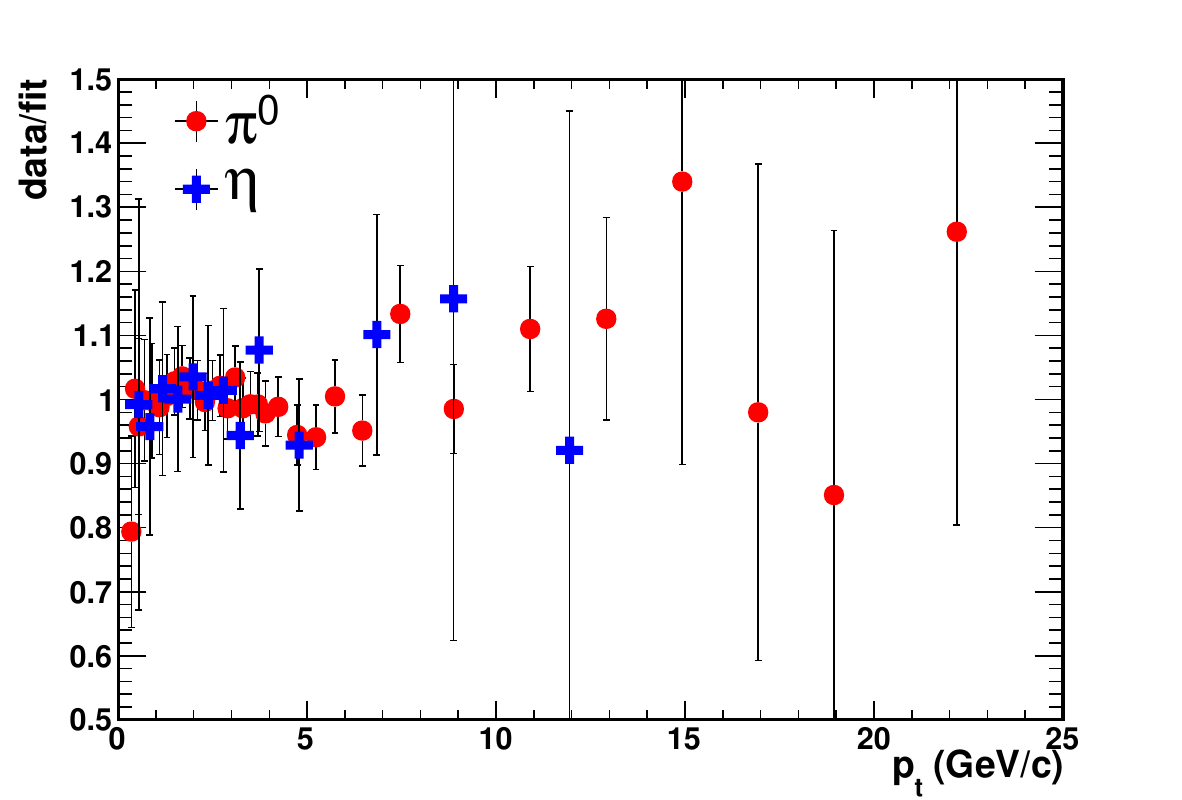}
\end{center}
\caption{(Colour online) Invariant differential production cross sections for
neutral pions and $\eta$ mesons in \pp collision at \mbox{$\sqrt{s} = 7$~TeV}
as function of \pt~\cite{Aamodt_pi0_eta} together with fits using
Eq.~\ref{fitfunc} (left panel). Ratios of the measured $\pi^0$ and $\eta$
spectra to the fits (right panel). In both panels the error bars depict the
combined statistical and systematic uncertainties of the neutral meson data.}
\label{fig:cocktail_input_mesons}
\end{figure}

\fi

The most important background source is the neutral pion. The contribution from 
$\pi^0$ decays to the background is twofold. First, the Dalitz decay of neutral
pions (\mbox{$\pi^0 \rightarrow$ e$^+$e$^-\gamma$}, with a branching ratio BR 
of $1.174 \pm 0.035\%$~\cite{Nakamura:2010zzi}) is a primary source of 
electrons from the collision vertex. Second, photons from the decay 
\mbox{$\pi^0 \rightarrow \gamma\gamma$} 
\mbox{(${\rm BR} = 98.823 \pm 0.034\%$~\cite{Nakamura:2010zzi})} can convert 
in material into \mbox{e$^{+}$e$^{-}$} pairs in the ALICE acceptance. 
This process gives rise to a secondary source of 
electrons not originating from the collision vertex. It is important to point 
out that, although the total material budget in the ALICE central barrel 
acceptance is relatively large ($11.4 \pm 0.5$\% of a radiation length 
$X_0$ integrated over a radial distance up to 180~cm from the beam line in the 
range \mbox{$|\eta| < 0.9$})~\cite{Koch:2011fw}, the material budget relevant 
for the present analysis is much less (see below).
In fact, electron candidate tracks considered here are 
required to be associated with either a hit in the first pixel layer of the 
ALICE ITS in case of the TPC-TOF/TPC-TRD-TOF analysis or a hit in any of the 
two pixel layers in the TPC-EMCal analysis. 
Therefore, only conversions in the beam pipe and in a fraction of the ITS 
material are relevant here. Consequently, the background contribution from 
photon conversions is similar to the contribution from Dalitz decays (see
below for a detailed calculation). 

The rapidity distribution of mesons is assumed to be flat around mid-rapidity. 
The momentum distributions of $\pi^0$ and $\eta$ mesons are obtained via 
fitting the spectra as measured by the ALICE 
collaboration~\cite{Aamodt_pi0_eta}. In this measurement, $\pi^0$ and $\eta$ 
decays in the $\gamma\gamma$ channel are reconstructed using two complementary 
techniques. As it is done conventionally, in the first approach the two decay 
photons are measured via electromagnetic calorimetry. This technique becomes 
notoriously difficult at low photon energy and, consequently, low meson \pt. 
In this region, it becomes advantageous to reconstruct photons in a second 
approach via the conversion into e$^{+}$e$^{-}$ pairs in the detector
acceptance. The large acceptance, high resolution ALICE TPC is ideally suited 
to perform such a measurement, which extends the $\pi^0$ spectrum down to 
300~MeV/$c$. Combining the measurements via calorimetry and the 
reconstruction of photon conversions, the $\pi^0$ and $\eta$ transverse 
momentum spectra from \pp collisions at \mbox{$\sqrt{s} = 7$~TeV} were 
measured by ALICE over a wide \pt range~\cite{Aamodt_pi0_eta}.

\ifrevtex
\input{tableTsallis_revtex.tex}
\else
\begin{table}
\caption{Fit parameters of the Tsallis parameterisation (see Eq.~\ref{fitfunc})
of the differential cross section of $\pi^0$ and $\eta$ meson production.}
\label{tab:tsallis_param}
\begin{center}
\begin{tabular}{c|c|c|c}
\hline\hline
Meson & \dndy & T (MeV) & n \\
\hline
$\pi^0$ & $2.40 \pm 0.15$ & $139 \pm 4$  & $6.88 \pm 0.07$ \\
$\eta$  & $0.21 \pm 0.03$ & $229 \pm 21$ & $7.0 \pm 0.5$ \\
\hline\hline
\end{tabular}
\end{center}
\end{table}

\fi

The invariant differential cross section of $\pi^0$ and $\eta$ meson production
in \pp collisions at \mbox{$\sqrt{s} = 7$~GeV} was parameterised with a Tsallis 
function~\cite{Tsallis:1987eu} given by:
\begin{equation}
E \frac{{\rm d}^3\sigma}{{\rm d}p^3} = 
\frac{\sigma_{\rm pp}}{2\pi}\frac{{\rm d}N}{{\rm d}y}\frac{(n-1)(n-2)}{nT(nT+m(n-2))}
(1+(m_{\rm t}-m)/(nT))^{-n},
\label{fitfunc}
\end{equation}
where the parameters d$N/{d\rm }y$, $T$, and $n$ were obtained by fitting the
experimental data as shown in Fig.~\ref{fig:cocktail_input_mesons}, 
$\sigma_{\rm pp}$ is the inelastic \pp cross section, $m$ is the relevant 
meson's mass and \mt is the corresponding transverse mass
$\mt = \sqrt{m^2 + \pt^2}$.
The values of the fit parameters are listed in Table~\ref{tab:tsallis_param}.

Given that pion decays and the corresponding conversion of decay photons are 
the most important cocktail ingredient up to intermediate \pt, the systematic 
uncertainty of the background cocktail is dominated by the uncertainty of the 
pion input. To evaluate this uncertainty the 
measured differential pion cross section was moved up (down) in all \pt
bins by the individual uncertainties in the bins, the parameterization
according to Eq.~\ref{fitfunc} was repeated, and full cocktails were generated
with these upper (lower) pion spectra as input. Thus, the uncertainty of the 
pion input was propagated to the electron cocktail. The same approach was 
followed for the $\eta$ meson.

Other light mesons ($\rho$, $\omega$, $\eta'$, and $\phi$) contribute
to the background electron cocktail through their Dalitz and/or dielectron 
decay channels as well as through the conversion of photons from their decays.
However, none of the contributions from these mesons is of any practical 
importance compared to the pion and the $\eta$ meson.
For the cocktail calculation, the shape of the invariant \pt distributions 
and the relative normalisations to the $\pi^0$ are required as input parameters
for the heavier mesons. The shape of the \pt spectra was derived from the pion
spectrum assuming \mt scaling, {\it i.e.} the spectral shapes of heavier
mesons and pions were consistent as a function of \mt.
Since the \mt scaling approach ensures that, at high \pt, the spectral shapes
of all meson distributions are the same, the normalisation of the heavier meson
spectra relative to the pion spectrum was determined by the ratios of heavier
meson yields to neutral pion yields at high \pt (\mbox{5~\gevc} in the present
analysis). The values used are summarised in Table~\ref{tab:meson_to_pion}. 
The quoted systematic uncertainties correspond to conservative estimates of 
30\% on all meson-to-pion ratios, which were propagated to the corresponding 
contributions to the background electron spectrum.

\ifrevtex
\input{tableCocktailsrc_revtex.tex}
\else
\begin{table}
\caption{Ratios of meson yields to neutral pion yields at \mbox{$\pt = 5$~\gevc}
in \pp~collisions at \mbox{$\sqrt{s} = 7$~TeV}.}
\label{tab:meson_to_pion}
\begin{center}
\begin{tabular}{c}
\hline\hline
$\rho / \pi^0 = 1.0 \pm 0.3$      \cite{Nakamura:2010zzi} \\
$\omega / \pi^0 = 0.85 \pm 0.255$ \cite{Nakamura:2010zzi,Adare:2011ht} \\
$\eta'  / \pi^0 = 0.25 \pm 0.075$ \cite{Nakamura:2010zzi} \\
$\phi / \pi^0 = 0.40 \pm 0.12$    \cite{Nakamura:2010zzi,Riabov:2007sq} \\
\hline\hline
\end{tabular}
\end{center}
\end{table}
\fi

A precise knowledge of the material budget is important for the calculation
of the electron spectrum from photon conversions.
An analysis of the reconstruction of photon conversions in ALICE demonstrated 
that the material budget implemented in the Monte Carlo simulation is in 
agreement within an uncertainty of 4.5\% with the actual material budget of the 
experiment~\cite{Koch:2011fw}. Since, for the present analysis, electron 
candidate tracks were required to be associated with a hit in the SPD, only the 
beam pipe, air, and a fraction of the ITS material contributed to the effective 
converter thickness. 
The beam pipe is made out of beryllium with a polyimide wrapping and its 
thickness in terms of radiation lengths is \mbox{$X/X_0 = 0.26\%$}. The 
corresponding thickness of a pixel layer is \mbox{$X/X_0 = 1.14\%$} for the 
full layer, including the sensor, the readout chip, and the 
infrastructure~\cite{ALICE}. 
The construction of the first pixel layer is such that the active sensor layer 
is closer to the beam line than the readout and most of the infrastructure, 
\ie conversions in the latter do not give rise to a recorded hit in this 
detector. In the second pixel layer, the order is reversed, \ie the readout 
and most of the infrastructure are closer to the beam line than the sensor 
itself. Therefore, for the TPC-EMCal analysis, the thickness of most of both 
pixel layers had to be considered in the calculation of the electron background
from photon conversions. Including an overall systematic uncertainty of 4.5\% 
on the material budget~\cite{Koch:2011fw}, the resulting converter thickness 
was \mbox{$X/X_0 = (2.15 \pm 0.11) \%$}, including the beam pipe and air, for 
photons impinging perpendicularly on the beam pipe and the ITS, \ie for 
photons at \mbox{$\eta = 0$}. For the TPC-TOF/TPC-TRD-TOF analysis only a 
fraction of the first pixel layer was relevant in addition to the beam pipe 
and air. For the latter case, from the known material budget and from full 
Monte Carlo simulations of photon conversions in the pixel detector the 
effective thickness of the first pixel layer was determined to be 
\mbox{$(45 \pm 5) \%$} of its total thickness. Including the beam pipe and air,
the effective converter thickness was \mbox{$X/X_0 = (0.77 \pm 0.07) \%$} at 
\mbox{$\eta = 0$}. The geometric $\eta$ dependence of the material budget was 
taken into account in the calculation of the photon conversion contribution in 
the electron background cocktail.

The ratio of conversion electrons to Dalitz electrons for $\pi^0$ decays was 
calculated as

\begin{equation}
\frac{\rm{Conversion}}{\rm{Dalitz}} = \frac{BR^{\gamma\gamma} \times 2 
                            \times (1 - e^{-\frac{7}{9}\times\frac{X}{X_0}}) 
                            \times 2}{BR^{\rm{Dalitz}} \times 2},
\label{Eq:conv_dalitz}
\end{equation}

where $BR^{\gamma\gamma}$ and $BR^{\rm{Dalitz}}$ are the branching ratios into the
two-photon and Dalitz channels, respectively. For the TPC-TOF/TPC-TRD-TOF 
analysis, with \mbox{$X/X_0 = (0.77 \pm 0.07) \%$}, this ratio 
$\rm{Conversion}/\rm{Dalitz}$ is equal to \mbox{$1.01 \pm 0.09$}. Due to the 
larger material budget relevant for the TPC-EMCal analysis, which is 
\mbox{$X/X_0 = (2.15 \pm 0.11) \%$}, the relative contribution from photon 
conversions to Dalitz decays was larger: 
$\rm{Conversion}/\rm{Dalitz} = 2.79 \pm 0.14$. 
For the decays of other light mesons the ratio is slightly smaller than for 
neutral pions due to the fact that \mbox{$BR^{\rm{Dalitz}}/BR^{\gamma\gamma}$} 
increases with increasing parent meson mass.

In addition, it was taken into account that the photon conversion probability 
is not constant but depends slightly on the photon energy, introducing a 
\pt dependence of the ratio $\rm{Conversion}/\rm{Dalitz}$, which was
determined in a full Monte Carlo simulation. The corresponding correction was 
applied in the calculation of the conversion contribution to the background 
electron cocktail. However, this correction is significant only for low
momentum electrons (\mbox{$0.5 < \pt < 1$~\gevc}), where the ratio 
$\rm{Conversion}/\rm{Dalitz}$ is reduced by 10\% or less relative to its
asymptotic value given in Eq.~\ref{Eq:conv_dalitz}.

The contribution from weak K$_{\rm e3}$ decays of charged and neutral kaons
can only be determined via simulations, which take into account the 
geometry of the experimental apparatus, the reconstruction algorithms, and 
the electron identification cuts. It turned out that the contribution 
from K$_{\rm e3}$ decays to the inclusive electron spectrum was essentially 
negligible. This was due to the fact that electron candidates considered in 
the present analysis were required to be associated with a hit in the first 
pixel layer of the ALICE ITS. Since this detector layer is close to the 
primary collision vertex (\mbox{3.9~cm} radial distance from the beam line) 
and because of the rather long life time of the relevant kaons 
(\mbox{$c\tau($K$^{\pm}) = 3.712$~m}, 
\mbox{$c\tau($K$^{0}_{\rm L}) = 15.34$~m}~\cite{Nakamura:2010zzi}), 
only a tiny fraction of K$_{\rm e3}$ decays contributed to the background 
electron sample. For electrons with \mbox{$\pt = 0.5$~\gevc} the relative 
contribution from K$_{\rm e3}$ decays to the inclusive electron background was 
not more than 0.5\%. For \mbox{$\pt = 1$~\gevc} this contribution decreased to 
\mbox{$\approx$0.2\%} and towards higher \pt it became even less. Given the 
limited statistics available in this simulation a conservative systematic 
uncertainty of 100\% is assigned to the K$_{\rm e3}$ contribution.

Electrons from the electromagnetic decays of heavy quarkonia have been added 
to the background electron cocktail based on measurements at the LHC. 
\mbox{J$/\psi$} production has been measured at mid-rapidity in \pp collisions 
at \mbox{7~TeV} by the ALICE~\cite{Aamodt:2011gj,jpsi_erratum} and CMS 
experiments~\cite{Khachatryan:2010yr}. 
A parameterisation of these data, obtained by a simultaneous fit according to 
Eq.~\ref{fitfunc} was used as input for the cocktail generator. 
$\Upsilon$ production at mid-rapidity has been measured by the CMS 
experiment~\cite{Khachatryan:2010zg}. As for the J$/\psi$, the production 
cross section was parametrised and the decay contribution was included in 
the electron cocktail.
While the contribution from J$/\psi$ decays becomes relevant at high \pt, the 
$\Upsilon$ contribution is negligible for the electron cocktail in the 
current \pt range.
While the systematic uncertainties of the measured production cross sections of 
heavy quarkonia were directly propagated to the corresponding decay electron 
spectra, their contribution to the systematic uncertainty of the latter is
less than 1\%.

Contributions to the background electron cocktail from prompt photons are 
twofold. Real photons produced in initial hard scattering processes, \eg via
quark-gluon Compton scattering, can convert in the detector material just as 
photons from meson decays. In addition, every source of real photons also 
emit virtual photons, \ie electron-positron pairs. The spectrum of real prompt
photons from an NLO pQCD 
calculation~\cite{Gordon:1993qc,Gordon:1994ut,Vogelsang:private} using
CTEQ6M5 parton distribution functions~\cite{Tung:2006tb} with GRV parton to
photon fragmentation functions~\cite{Gluck:1992zx,Gluck:1992zxe} was 
parameterised, and the corresponding conversion electron spectrum was added to 
the background electron cocktail. The ratio of virtual prompt photons to real 
prompt photons increases with increasing \pt because the phase space for 
dielectron emission increases~\cite{PhysRevC.81.034911}. This has been taken 
into account in the calculation of the corresponding contribution to the 
background electron cocktail.
Prompt photon production has not been measured in ALICE yet. Measurements at 
lower collision energy are in agreement with NLO pQCD calculations within 
uncertainties of significantly less than 50\% at high \pt~\cite{Adler:2006yt}. 
Conservatively, a systematic uncertainty of 50\% was assigned to the 
contribution from prompt photons to the total background electron cocktail.

Contributions from the Drell-Yan process are expected to be small in the
\pt range covered by the present analysis and, therefore, they were not included in the background electron cocktail.

\ifrevtex
\input{tableCocktailsys_revtex.tex}
\else
\begin{table}
\begin{center}
\caption{Overview over the contributions to the systematic uncertainties of the
background cocktail. The contributions from mesons heavier than the $\eta$ 
meson and the contribution from $K_{\rm e3}$ decays to the systematic uncertainty
are less than 1\% and, therefore, are not listed explicitly. For details on 
the error determination, see text.}
\label{sys_cocktail}
\begin{tabular}{l|c|c|c}
\hline\hline
\pt (\gevc) & 0.5 & 2 & 8\\
\hline
Error source & \multicolumn{3}{c}{systematic uncertainty (\%)}\\
\hline
$\pi^0$ spectrum & $\pm8$ & $\pm4$ & $\pm8$ \\
$\gamma$ conversions & $\pm4$ & $\pm4$ & $\pm3$ \\
$\eta$ spectrum & $\pm1$ & $\pm1$ & $\pm4$ \\
prompt photons & $<\pm1$ & $<\pm1$ & $\pm4$ \\
\hline
total & $\pm9$ & $\pm6$ & $\pm10$ \\
\hline\hline
\end{tabular}
\end{center}
\end{table}

\fi

To calculate the systematic uncertainty of the cocktails, the 
systematic uncertainties of all uncorrelated cocktail ingredients were 
estimated as discussed above, propagated to the corresponding electron spectra,
and added in quadrature. The cocktail systematic uncertainties are smallest in 
the \pt range between 1 and 2~\gevc. The individual contributions and their 
dependence on \pt are summarised in Table~\ref{sys_cocktail}, where error 
sources with less than 1\% systematic uncertainty are not listed.

The total background cocktail electron cross sections were divided by the 
minimum bias \pp cross section 
\mbox{$62.2 \pm 2.2 ({\rm sys.})$~mb}~\cite{MBcrosssection} 
(see below) such that they can be directly compared to the measured inclusive 
electron yields per minimum bias triggered collision. These comparisons are 
shown in Fig.~\ref{fig:incl_vs_cocktail} for the TPC-TOF/TPC-TRD-TOF analysis 
(left panel) and the TPC-EMCal analysis (right panel).

\section{Results and discussion}
\label{sec:results}
\ifrevtex
\input{figInclusiveCocktail_revtex.tex}
\else
\begin{figure}[tbh]
\begin{center}
\includegraphics[width=0.495\linewidth]{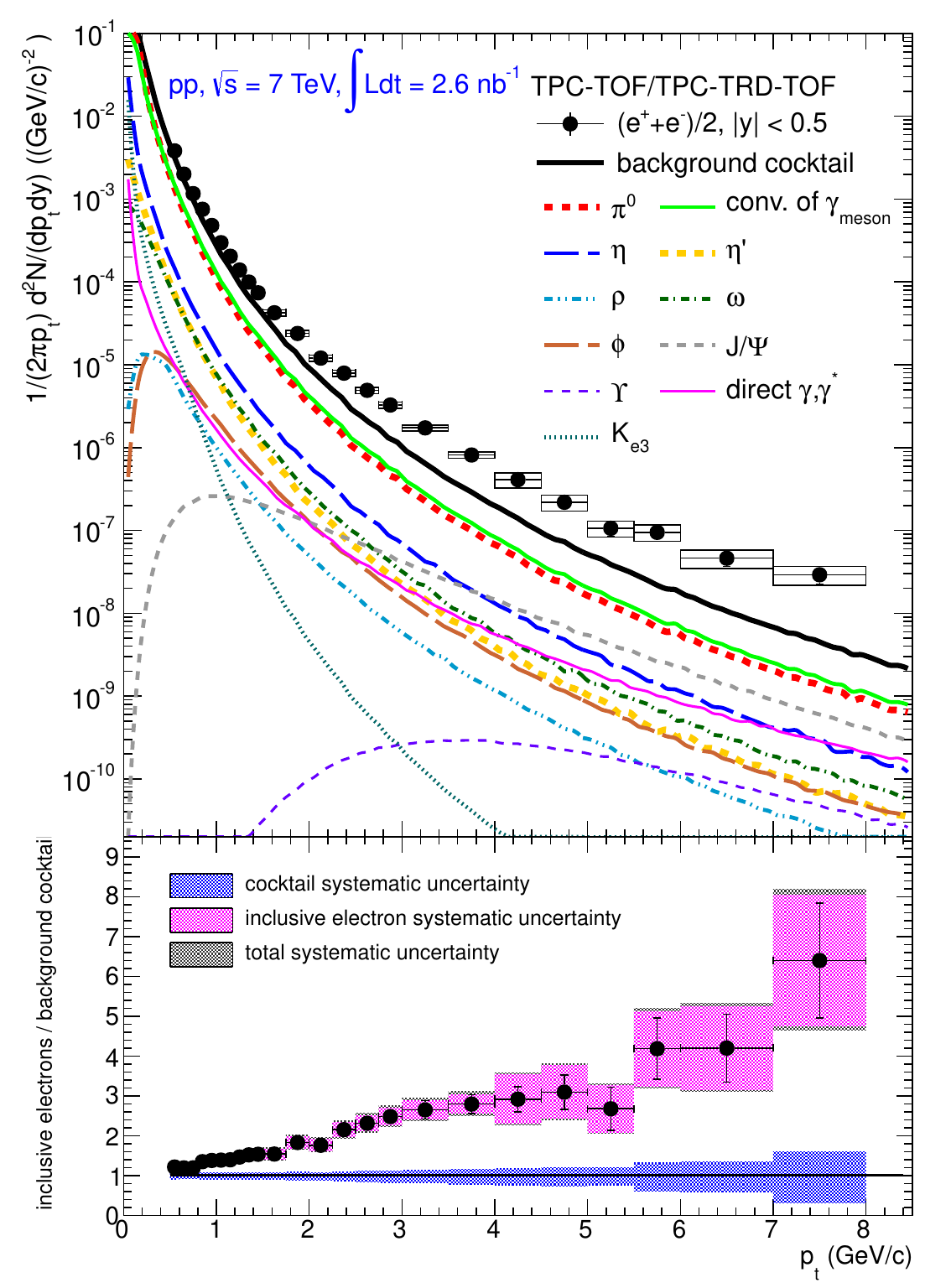}
\includegraphics[width=0.495\linewidth]{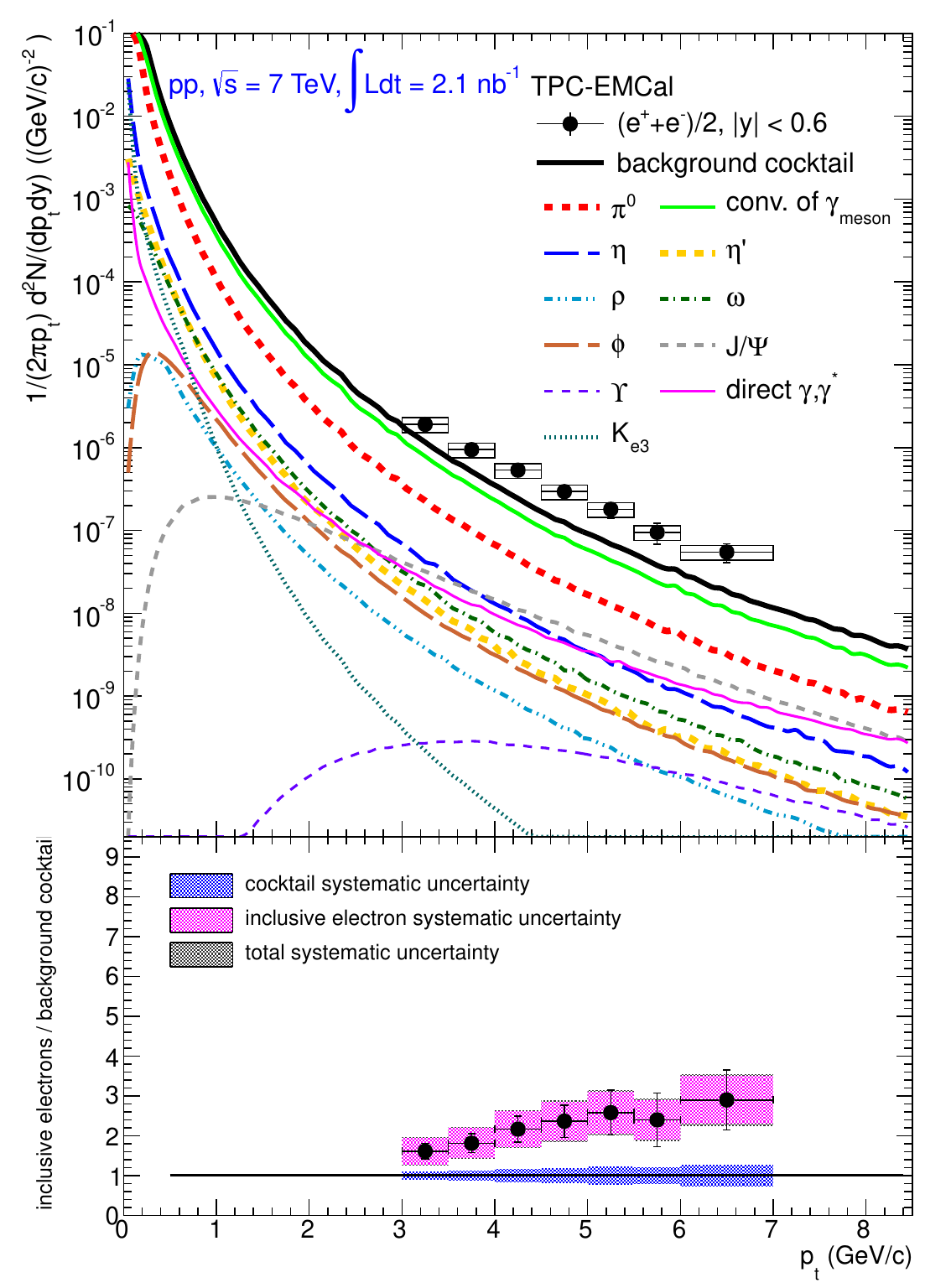}
\end{center}
\caption{(Colour online) Inclusive electron yield per minimum bias \pp
collision as function of \pt at \mbox{\s$ = 7$~TeV} in comparison with
background electron cocktails for the TPC-TOF/TPC-TRD-TOF analysis (left panel)
and the TPC-EMCal analysis (right panel). Lower panels show the ratio of the
inclusive electron yield to the background electron cocktail for both analyses.}
\label{fig:incl_vs_cocktail}
\end{figure}

\fi
\subsection{Heavy-flavour hadron decay electron cross section}
\label{subsec:hfe}

The differential inclusive electron yield in \pp collisions at 
\mbox{\s$ = 7$~TeV}, already shown in Fig.~\ref{fig:ince}, is compared to the 
background electron yield as calculated within the cocktail approach in the 
left and right panels of Fig.~\ref{fig:incl_vs_cocktail} for the 
TPC-TOF/TPC-TRD-TOF and the TPC-EMCal analysis, respectively. 
Statistical uncertainties in the
inclusive electron measurement are shown as error bars, while systematic
uncertainties are indicated by boxes. The background contribution from
photon conversions is smaller in the TPC-TOF/TPC-TRD-TOF analysis because in 
this case a hit in the first pixel layer is required for electron candidate 
tracks. Consequently, the ratio of the measured inclusive electron yield
to the calculated electron background is larger for the TPC-TOF/TPC-TRD-TOF 
analysis than for the TPC-EMCal analysis as shown in the lower left and right
panels of Fig.~\ref{fig:incl_vs_cocktail}, respectively. 

The differential production cross section of electrons from heavy-flavour 
decays is calculated by first subtracting the background cocktail from the 
inclusive electron spectrum and then multiplying the difference with the
minimum bias \pp cross section $\sigma_{\rm MB}$. The corresponding systematic 
uncertainties propagated from the inclusive electron measurement and the 
electron background cocktail are summarised in Table~\ref{sys_hfe}.
The value for $\sigma_{\rm MB}$ is \mbox{$62.2 \pm 2.2 ({\rm sys.})$~mb}. 
This number was obtained by relating $\sigma_{\rm MB}$ to the cross section 
$\sigma_{\rm VOAND}$ sampled with the V0AND trigger~\cite{MBcrosssection}.  
The latter corresponds to the coincidence between signals in the two VZERO 
detectors as measured in a van der Meer scan~\cite{VanderMeer}. The relative 
factor $\sigma_{\rm VOAND} / \sigma_{\rm MB}$ is equal to 0.873 and stable within 
1\% over the analysed data sample. The corresponding systematic uncertainty of 
3.5\% is due to uncertainties of the measured beam intensities and in the 
analysis procedure of the van der Meer scan~\cite{SysV0AND}. As demonstrated in 
Fig.~\ref{fig:hfe}, the resulting cross sections from the TPC-TOF/TPC-TRD-TOF 
and TPC-EMCal analyses agree with each other within the experimental 
uncertainties.

\ifrevtex
\input{tableHFESysError_revtex.tex}
\else
\begin{table}
\begin{center}
\caption{Systematic uncertainties of the electron cross section from 
heavy-flavour hadron decays propagated from the inclusive electron measurement
and the electron background cocktail for the TPC-TOF/TPC-TRD-TOF analysis. For 
$\pt = 3$, 5, and 7~\gevc the corresponding uncertainties for the TPC-EMCal 
analysis are quoted in parentheses.}
\label{sys_hfe}
\begin{tabular}{l|c|c|c|c}
\hline\hline
\pt (\gevc) & 1 & 3 & 5 & 7 \\
\hline
Error source & \multicolumn{4}{c}{Systematic uncertainty (\%)}\\
\hline
& & & & \\
Inclusive electron spectrum  & $^{+32}_{-36}$ & $\pm 14 (\pm 53)$ & 
                               $\pm 33 (\pm 35)$ & 
                               $\pm 31 (\pm 31)$ \\
& & & & \\
Electron background cocktail & $^{+17}_{-13}$ & $^{+5}_{-3} (^{+10}_{-11})$ & 
                               $^{+5}_{-4} (\pm 5)$ & 
                               $^{+3}_{-2} (\pm 5)$ \\
& & & & \\
\hline
Total & $^{+36}_{-38}$ & $^{+15}_{-14} (\pm 54)$ & $^{+34}_{-33} (\pm 35)$ & $\pm 31 (\pm 31)$ \\
\hline\hline
\end{tabular}
\end{center}
\end{table}

\fi

Since the azimuthal coverages of the TRD and the EMCal are mutually exclusive 
and because the electron identification is done following different approaches,
the statistical uncertainties of the inclusive electron spectra measured, using 
these two methods, are uncorrelated.
While the systematic uncertainties related to the electron identification
are essentially uncorrelated, those originating from the track reconstruction
are mostly correlated. In addition, the systematic uncertainties of the 
electron background cocktails are correlated completely for both analyses. 

The final production cross section for electrons from heavy-flavour hadron 
decays is calculated as the weighted average of the TPC-TOF/TPC-TRD-TOF and 
TPC-EMCal measurements, where the weights are calculated from the quadratic 
sums of the statistical and uncorrelated systematic uncertainties of the 
individual analyses.
To determine the uncertainties of the weighted average, uncorrelated 
uncertainties of the two analyses are added in quadrature while correlated
uncertainties are added linearly. 

\ifrevtex
\input{figHFEcomparison_revtex.tex}
\else
\begin{figure}[tbh]
\begin{center}
\includegraphics[width=0.8\linewidth]{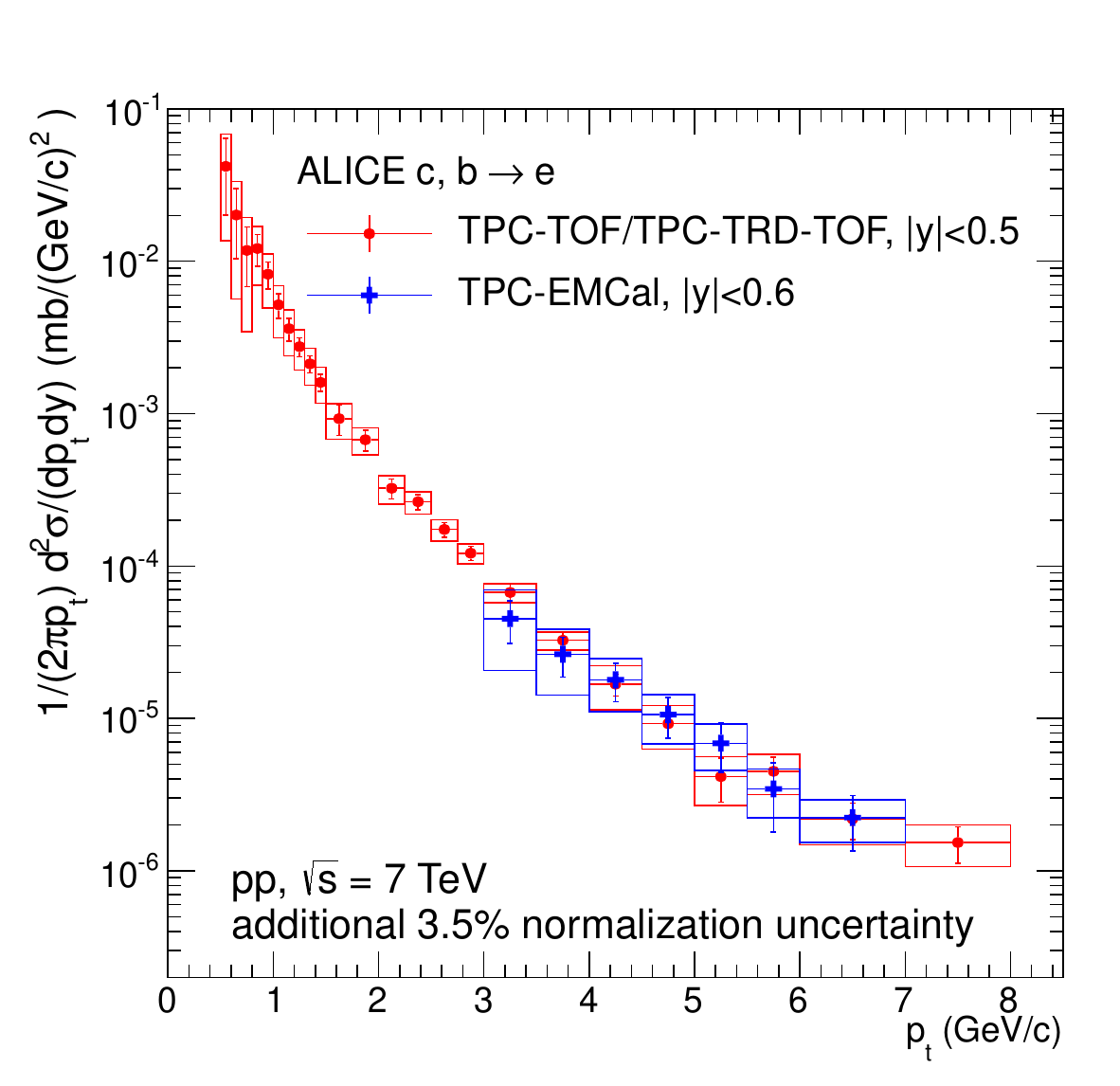}
\end{center}
\caption{(Colour online) Invariant differential production cross section for
electrons from heavy-flavour decays in \pp collisions at \mbox{\s$ = 7$~TeV}
for the TPC-TOF/TPC-TRD-TOF and the TPC-EMCal measurements. The overall
systematic uncertainty of 3.5\% on the cross section normalisation is not
included.}
\label{fig:hfe}
\end{figure}

\fi

The differential invariant cross section of electrons from semileptonic 
heavy-flavour decays is measured for transverse momenta above 0.5~\gevc. 
It is interesting to note that according to calculations using the 
PYTHIA 6.4.21 event generator~\cite{Sjostrand:2006za} with the Perugia-0 
parameter tuning~\cite{Skands:2009zm} $\sim 57$\% of the electrons from charm 
decays and $\sim 73$\% of the electrons from beauty decays are within the 
measured \pt range in the rapidity interval $|y| < 0.5$. For 
FONLL~\cite{fonll,fonll2,fonll3} pQCD 
calculations similar values are obtained. In this case, $\sim 51$\% of the 
electrons from charm decays and $\sim 90$\% of the electrons from beauty 
decays are within the accessible \pt range.

\ifrevtex
\input{figHFE_vs_pizero_revtex.tex}
\else
\begin{figure}[tbh]
\begin{center}
\includegraphics[width=0.8\linewidth]{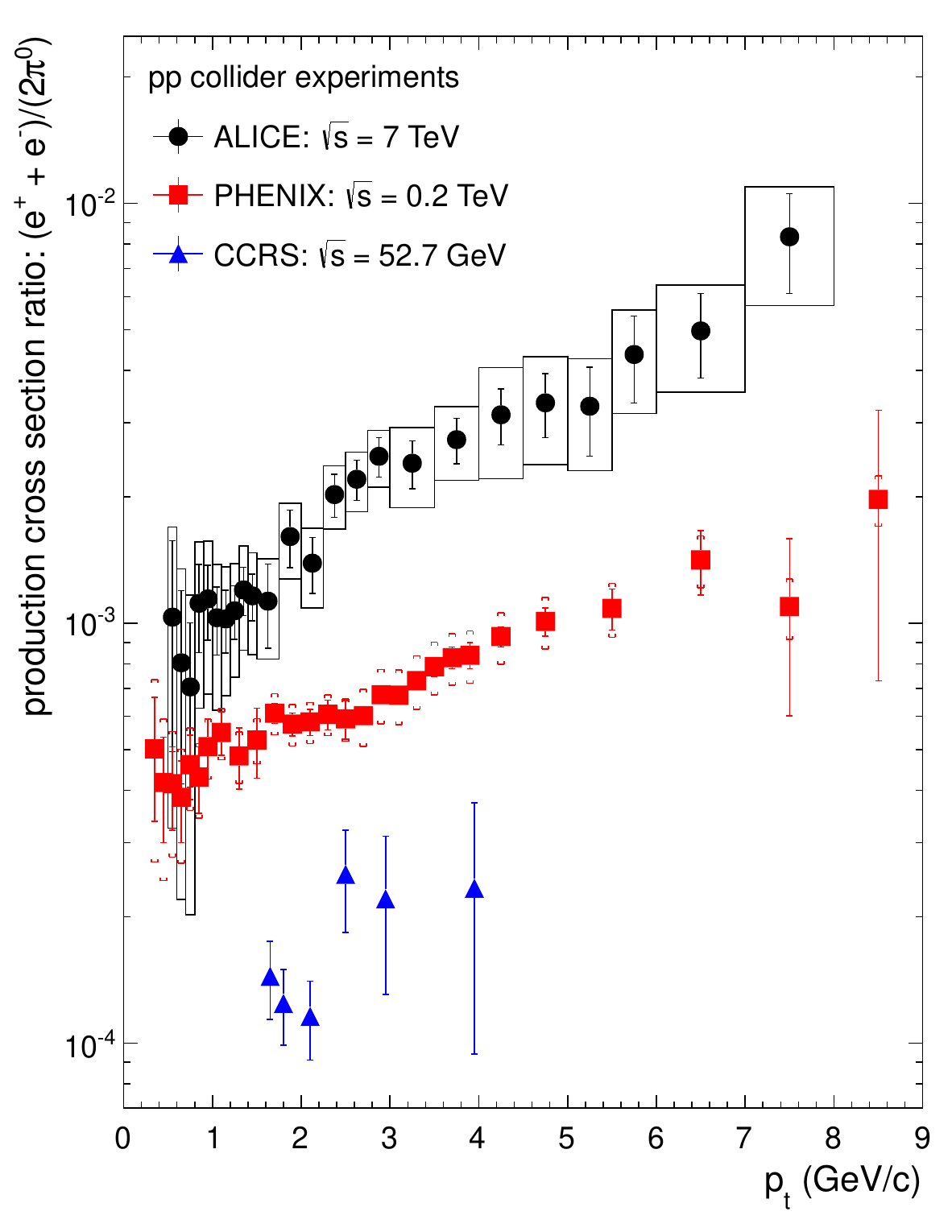}
\end{center}
\caption{(Colour online) Production cross section ratios of electrons from
heavy-flavour hadron decays and neutral pions in \pp collisions measured
as function of \pt with ALICE at the LHC, PHENIX at RHIC~\cite{Adare:2010de,PhysRevLett.91.072301,PhysRevLett.101.232301,PhysRevC.69.034909}, and the CCRS
experiment at the ISR~\cite{Busser:1974ej,Alper:1975}.
Error bars are statistical errors. Systematic uncertainties are depicted by
boxes (ALICE) and brackets (PHENIX).}
\label{fig:hfe_vs_pizero}
\end{figure}

\fi

The ratio \mbox{(e$^+$+e$^-$)/2$\pi^0$} of the production cross sections of 
electrons from heavy-flavour hadron decays and $\pi^0$ mesons, which are the 
main source of background in the relevant \pt range, is depicted as a function 
of \pt in Fig.~\ref{fig:hfe_vs_pizero}. 
The comparison with corresponding measurements from lower energy \pp collider 
experiments demonstrates the different $\sqrt{s}$ dependence of the heavy and 
light flavour production cross sections.
At the CERN ISR, the CCRS experiment recorded a first low statistics data 
sample of electrons from heavy-flavour hadron decays in \pp collisions at 
$\sqrt{s} = 52.7$~GeV in the range $1.6 < \pt < 4.7$~\gevc~\cite{Busser:1974ej}.
The charge-averaged inclusive charged pion production cross section was 
measured and parameterised by the British-Scandinavian Collaboration in the 
same collision system~\cite{Alper:1975}. 
Assuming that the latter is equal to the neutral pion production cross section
the ratio \mbox{(e$^+$+e$^-$)/2$\pi^0$} was calculated with substantial 
statistical uncertainties as shown in Fig.~\ref{fig:hfe_vs_pizero}. 
At ISR energies, the ratio \mbox{(e$^+$+e$^-$)/2$\pi^0$} is of the order 
$(1-2) \times 10^{-4}$ without a significant \pt dependence in the range 
below 5~\gevc, which is dominated by charm hadron decays.
At $\sqrt{s} = 0.2$~TeV, production cross sections of both electrons from
heavy-flavour hadron decays~\cite{Adare:2010de} and 
pions~\cite{PhysRevLett.91.072301,PhysRevLett.101.232301,PhysRevC.69.034909} 
have been measured with the PHENIX experiment in \pp collisions at RHIC. 
The ratio \mbox{(e$^+$+e$^-$)/2$\pi^0$} was evaluated at given values of \pt in
the range $0.3 < \pt < 9$~\gevc as shown in Fig.~\ref{fig:hfe_vs_pizero}. 
In the charm-hadron decay dominated low \pt region, the larger ratio 
\mbox{(e$^+$+e$^-$)/2$\pi^0 \approx 5 \times 10^{-4}$} is consistent with a more
rapid increase of the charm production cross section with $\sqrt{s}$ as 
compared to the light-flavour production cross section.
At RHIC, a pronounced \pt dependence of the ratio \mbox{(e$^+$+e$^-$)/2$\pi^0$}
is observed. From the low \pt ($\sim$1~\gevc) to the beauty-hadron decay 
dominated high \pt region ($\sim$9~\gevc) \mbox{(e$^+$+e$^-$)/2$\pi^0$} rises 
by at least a factor two, consistent with the beauty production cross section 
rising faster with $\sqrt{s}$ than the charm production cross section. 
At the LHC, the increase of the ratio \mbox{(e$^+$+e$^-$)/2$\pi^0$} with \pt is
even larger. In the present measurement at $\sqrt{s} = 7$~TeV, 
\mbox{(e$^+$+e$^-$)/2$\pi^0$} grows by almost an order of magnitude from the 
charm decay dominated low \pt region to $\approx 10^{-2}$ in the beauty 
dominated high \pt region. 

\subsection{Comparison with FONLL pQCD}
\label{subsec:fonll}
The measured differential invariant production cross section of electrons 
from heavy-flavour decays is compared with a FONLL pQCD calculation in 
Fig.~\ref{fig:hfe_vs_fonll}, where error bars depict the statistical 
uncertainty while boxes show the total systematic uncertainty of the 
measurement. For the FONLL calculation CTEQ6.6 parton distribution 
functions~\cite{CTEQ66} were used. To obtain the uncertainty of the 
calculation, indicated by dashed-dotted lines in Fig.~\ref{fig:hfe_vs_fonll},
the factorisation and renormalisation scales $\mu_{\rm F}$ 
and $\mu_{\rm R}$, respectively, were varied independently in the ranges 
\mbox{$0.5 < \mu_{\rm F}/m_{\rm t} < 2$} and 
\mbox{$0.5 < \mu_{\rm R}/m_{\rm t} < 2$}, with the additional constraint 
\mbox{$0.5 < \mu_{\rm F}/\mu_{\rm R} < 2$}, where \mt is the transverse mass
of the heavy quarks. 
The charm quark mass was varied in FONLL within the range 
\mbox{$1.3 < m_{\rm c} < 1.7$~GeV/$c^{2}$} and the beauty quark mass was varied 
within \mbox{$4.5 < m_{\rm b} < 5.0$~GeV/$c^{2}$}~\cite{fonll}. For electrons
from charm hadron decays, the contributions from D$^{0}$ and D$^{+}$ decays
were weighted with the measured D$^{0}$/D$^{+}$ ratio \cite{ALICE-D2H}.
Variations due to different choices of the parton distributions functions
were also included in the theoretical uncertainty. The differential cross 
section of electrons from heavy-flavour decays in the rapidity interval 
\mbox{$|y| < 0.5$} is shown in comparison with the FONLL prediction on an 
absolute scale in the upper panel of Fig.~\ref{fig:hfe_vs_fonll}. In addition 
to charm and beauty hadron decays to electrons also the cascade beauty to 
charm to electron is included. Statistical and systematic uncertainties of the 
measurement are depicted as error bars and boxes, respectively. The cross 
section and uncertainty from FONLL are shown as solid and dashed-dotted lines,
respectively. 

The ratio of the measured cross section and the FONLL calculation
is drawn in the lower panel of Fig.~\ref{fig:hfe_vs_fonll}. Error bars and 
boxes around the data points indicate the statistical and systematic
uncertainties of the electron spectrum from heavy-flavour decays, respectively.
These systematic error boxes do not include any contribution from the FONLL 
calculation. The relative systematic uncertainties of the plotted ratio 
originating from the FONLL calculation is indicated by the dashed-dotted lines
around one. Within substantial theoretical uncertainties the FONLL pQCD 
calculation is in agreement with the data.           

\ifrevtex
\input{figComparisonFONLL_revtex.tex}
\else
\begin{figure}[tbh]
\begin{center}
\includegraphics[width=0.7\linewidth]{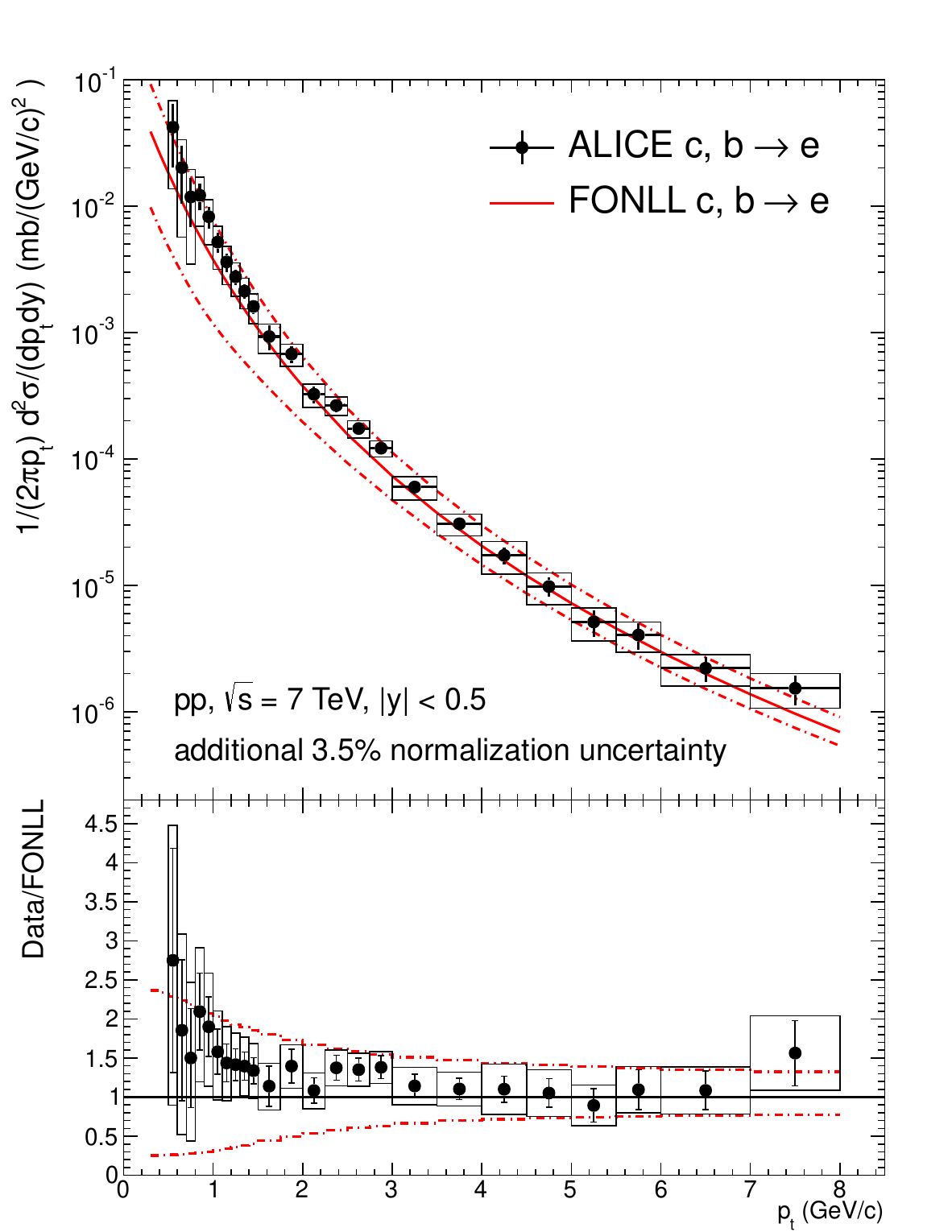}
\end{center}
\caption{(Colour online) The measured electron spectrum from heavy-flavour
hadron decays is compared to a FONLL calculation for inclusive charm and
beauty hadron semileptonic decays on an absolute scale in the upper panel.
The ratio of the measured spectrum to the FONLL pQCD calculation is shown in
the lower panel. Error bars, boxes, and theoretical uncertainties are
described in the text.}
\label{fig:hfe_vs_fonll}
\end{figure}

\fi

\subsection{ALICE and ATLAS measurements of electrons from heavy-flavour 
hadron decays}
\label{subsec:atlas}

The ATLAS experiment has measured electrons from heavy-flavour decays in \pp
collisions at \mbox{$\sqrt{s} = 7$~TeV} in the \pt range 
\mbox{$7 < \pt < 26$~\gevc} and in the rapidity interval $|y| < 2$, where the 
regions \mbox{$1.37 < |y| < 1.52$} are excluded~\cite{Aad:2011rr}.
The \pt-differential production cross section, $d\sigma/d\pt$, published by 
ATLAS is divided bin-by-bin by $2 \pi \pt \Delta y$, where $\pt$ is the center 
of the individual transverse momentum bins chosen by ATLAS and $\Delta y$ is 
the rapidity interval covered by the ATLAS measurement. The result is shown 
together with the electron cross section 
presented in this paper in Fig.~\ref{fig:hfe_vs_atlas}. While the electron 
measurement by ALICE includes most of the total cross section, the data from 
ATLAS extend the measurement to higher \pt. 
Corresponding FONLL pQCD calculations in the rapidity intervals covered by
ALICE and ATLAS, respectively, are included for comparison in 
Fig.~\ref{fig:hfe_vs_atlas} as well. Within the experimental and theoretical
uncertainties FONLL is in agreement with both data sets.
It should be noted that the invariant cross section per unit rapidity
decreases with increasing width of the rapidity interval because the 
heavy-flavour production cross section decreases towards larger absolute 
rapidity values. However, this effect is small in \pp collisions at 
$\sqrt{s} = 7$~TeV ($< 5\%$ for electrons from charm decays and $< 10\%$
for electrons from beauty decays according to FONLL calculations).

\ifrevtex
\input{figComparisonATLAS_revtex.tex}
\else
\begin{figure}[tbh]
\begin{center}
\includegraphics[width=0.7\linewidth]{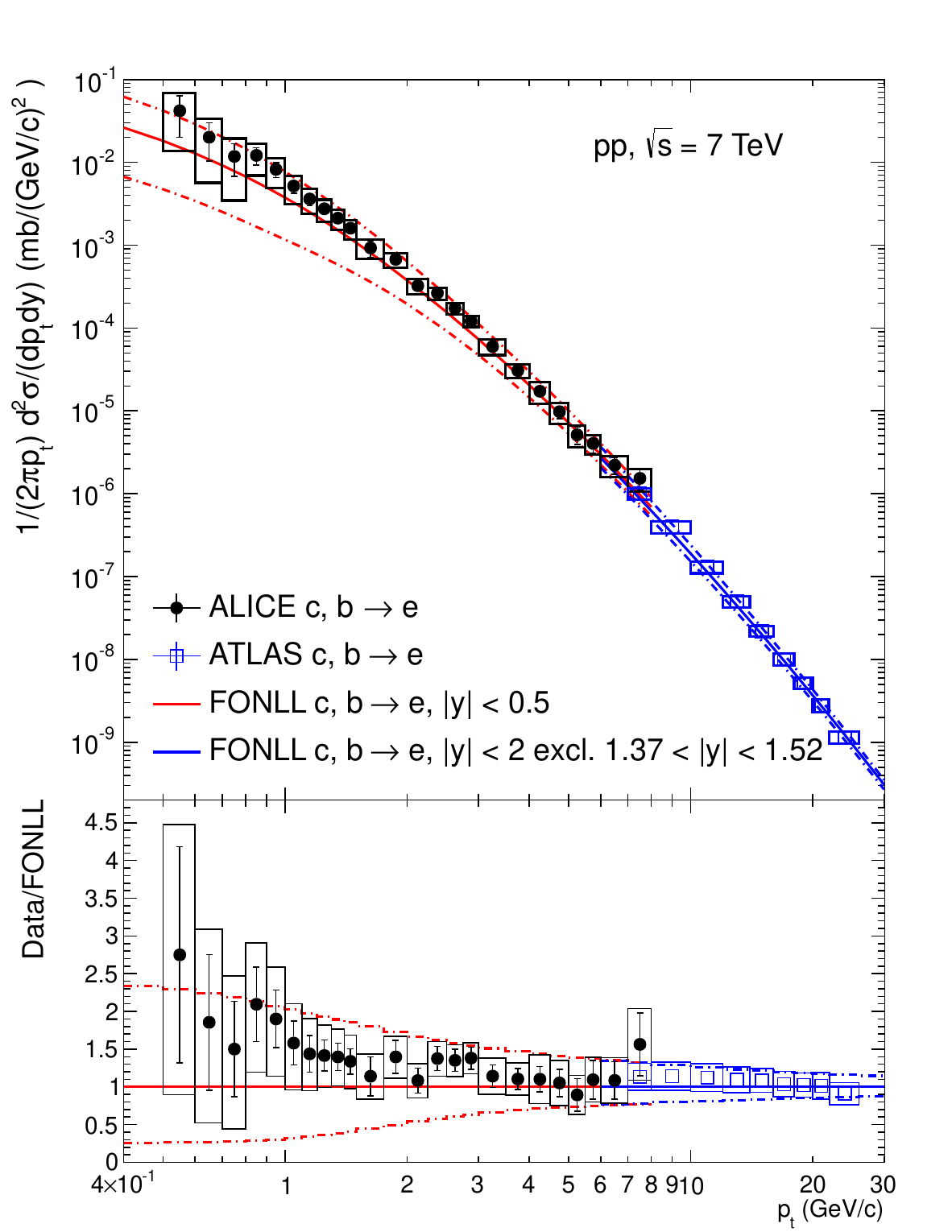}
\end{center}
\caption{(Colour online): Invariant differential production cross sections
of electrons from heavy-flavour decays measured by ALICE and
ATLAS~\cite{Aad:2011rr} in \pp collisions at $\sqrt{s} = 7$~TeV in different
rapidity intervals (see text). FONLL pQCD calculations with the same rapidity
selections are shown for comparison.}
\label{fig:hfe_vs_atlas}
\end{figure}

\fi

\section{Summary}
\label{sec:summary}
The inclusive differential production cross section of electrons from
charm and beauty decays has been measured by ALICE in the transverse
momentum range \mbox{$0.5 <$\pt$<8$~\gevc} at mid-rapidity in \pp
collisions at \mbox{\s$= 7$~TeV}. Within experimental and theoretical
uncertainties a perturbative QCD calculation in the framework of 
FONLL is consistent with the measured differential cross section.
The data presented in this paper extend a corresponding measurement from
ATLAS, which is restricted to the high \pt region, towards substantially
lower transverse momenta. This low \pt region includes the dominant fraction
of the total heavy-flavor production cross section, and future higher precision
data might be sensitive to the parton distribution function of the proton
at low $x$.

\section*{Acknowledgements}
\label{sec:acknowledgements_Jan2012}
The ALICE collaboration would like to thank all its engineers and technicians for their invaluable contributions to the construction of the experiment and the CERN accelerator teams for the outstanding performance of the LHC complex.
\\
The ALICE collaboration would like to thank M. Cacciari for providing the FONLL pQCD predictions for the cross sections of 
electrons from heavy-flavour hadron decays. Furthermore, the ALICE collaboration would like to thank W. Vogelsang for 
providing the NLO pQCD predictions for direct photon production cross sections which were used as one of the inputs 
for the electron background cocktail.
\\
The ALICE collaboration acknowledges the following funding agencies for their support in building and
running the ALICE detector:
 \\
Calouste Gulbenkian Foundation from Lisbon and Swiss Fonds Kidagan, Armenia;
 \\
Conselho Nacional de Desenvolvimento Cient\'{\i}fico e Tecnol\'{o}gico (CNPq), Financiadora de Estudos e Projetos (FINEP),
Funda\c{c}\~{a}o de Amparo \`{a} Pesquisa do Estado de S\~{a}o Paulo (FAPESP);
 \\
National Natural Science Foundation of China (NSFC), the Chinese Ministry of Education (CMOE)
and the Ministry of Science and Technology of China (MSTC);
 \\
Ministry of Education and Youth of the Czech Republic;
 \\
Danish Natural Science Research Council, the Carlsberg Foundation and the Danish National Research Foundation;
 \\
The European Research Council under the European Community's Seventh Framework Programme;
 \\
Helsinki Institute of Physics and the Academy of Finland;
 \\
French CNRS-IN2P3, the `Region Pays de Loire', `Region Alsace', `Region Auvergne' and CEA, France;
 \\
German BMBF and the Helmholtz Association;
\\
General Secretariat for Research and Technology, Ministry of
Development, Greece;
\\
Hungarian OTKA and National Office for Research and Technology (NKTH);
 \\
Department of Atomic Energy and Department of Science and Technology of the Government of India;
 \\
Istituto Nazionale di Fisica Nucleare (INFN) of Italy;
 \\
MEXT Grant-in-Aid for Specially Promoted Research, Ja\-pan;
 \\
Joint Institute for Nuclear Research, Dubna;
 \\
National Research Foundation of Korea (NRF);
 \\
CONACYT, DGAPA, M\'{e}xico, ALFA-EC and the HELEN Program (High-Energy physics Latin-American--European Network);
 \\
Stichting voor Fundamenteel Onderzoek der Materie (FOM) and the Nederlandse Organisatie voor Wetenschappelijk Onderzoek (NWO), Netherlands;
 \\
Research Council of Norway (NFR);
 \\
Polish Ministry of Science and Higher Education;
 \\
National Authority for Scientific Research - NASR (Autoritatea Na\c{t}ional\u{a} pentru Cercetare \c{S}tiin\c{t}ific\u{a} - ANCS);
 \\
Federal Agency of Science of the Ministry of Education and Science of Russian Federation, International Science and
Technology Center, Russian Academy of Sciences, Russian Federal Agency of Atomic Energy, Russian Federal Agency for Science and Innovations and CERN-INTAS;
 \\
Ministry of Education of Slovakia;
 \\
Department of Science and Technology, South Africa;
 \\
CIEMAT, EELA, Ministerio de Educaci\'{o}n y Ciencia of Spain, Xunta de Galicia (Conseller\'{\i}a de Educaci\'{o}n),
CEA\-DEN, Cubaenerg\'{\i}a, Cuba, and IAEA (International Atomic Energy Agency);
 \\
Swedish Research Council (VR) and Knut $\&$ Alice Wallenberg
Foundation (KAW);
 \\
Ukraine Ministry of Education and Science;
 \\
United Kingdom Science and Technology Facilities Council (STFC);
 \\
The United States Department of Energy, the United States National
Science Foundation, the State of Texas, and the State of Ohio.

\bibliographystyle{unsrt}
\bibliography{hfe_pp}

\newpage
\begin{appendix}
\section{The ALICE Collaboration}
\label{authorlist}

\begingroup
\small
\begin{flushleft}
B.~Abelev\Irefn{org1234}\And
J.~Adam\Irefn{org1274}\And
D.~Adamov\'{a}\Irefn{org1283}\And
A.M.~Adare\Irefn{org1260}\And
M.M.~Aggarwal\Irefn{org1157}\And
G.~Aglieri~Rinella\Irefn{org1192}\And
A.G.~Agocs\Irefn{org1143}\And
A.~Agostinelli\Irefn{org1132}\And
S.~Aguilar~Salazar\Irefn{org1247}\And
Z.~Ahammed\Irefn{org1225}
% ...
\renewcommand\labelenumi{\textsuperscript{\theenumi}~}
\section*{Affiliation notes}
\renewcommand\theenumi{\roman{enumi}}
\begin{Authlist}
\item \Adef{M.V.Lomonosov Moscow State University, D.V.Skobeltsyn Institute of Nuclear Physics, Moscow, Russia}Also at: M.V.Lomonosov Moscow State University, D.V.Skobeltsyn Institute of Nuclear Physics, Moscow, Russia
\item \Adef{Institute of Nuclear Sciences, Belgrade, Serbia}Also at: "Vin\v{c}a" Institute of Nuclear Sciences, Belgrade, Serbia
\end{Authlist}
\section*{Collaboration Institutes}
\renewcommand\theenumi{\arabic{enumi}~}
\begin{Authlist}
\item \Idef{org1279}Benem\'{e}rita Universidad Aut\'{o}noma de Puebla, Puebla, Mexico
\item \Idef{org1220}Bogolyubov Institute for Theoretical Physics, Kiev, Ukraine
\item \Idef{org1262}Budker Institute for Nuclear Physics, Novosibirsk, Russia
\item \Idef{org1292}California Polytechnic State University, San Luis Obispo, California, United States
\item \Idef{org14939}Centre de Calcul de l'IN2P3, Villeurbanne, France
\item \Idef{org1197}Centro de Aplicaciones Tecnol\'{o}gicas y Desarrollo Nuclear (CEADEN), Havana, Cuba
\item \Idef{org1242}Centro de Investigaciones Energ\'{e}ticas Medioambientales y Tecnol\'{o}gicas (CIEMAT), Madrid, Spain
\item \Idef{org1244}Centro de Investigaci\'{o}n y de Estudios Avanzados (CINVESTAV), Mexico City and M\'{e}rida, Mexico
\item \Idef{org1335}Centro Fermi -- Centro Studi e Ricerche e Museo Storico della Fisica ``Enrico Fermi'', Rome, Italy
\item \Idef{org17347}Chicago State University, Chicago, United States
\item \Idef{org1288}Commissariat \`{a} l'Energie Atomique, IRFU, Saclay, France
\item \Idef{org1294}Departamento de F\'{\i}sica de Part\'{\i}culas and IGFAE, Universidad de Santiago de Compostela, Santiago de Compostela, Spain
\item \Idef{org1106}Department of Physics Aligarh Muslim University, Aligarh, India
\item \Idef{org1121}Department of Physics and Technology, University of Bergen, Bergen, Norway
\item \Idef{org1162}Department of Physics, Ohio State University, Columbus, Ohio, United States
\item \Idef{org1300}Department of Physics, Sejong University, Seoul, South Korea
\item \Idef{org1268}Department of Physics, University of Oslo, Oslo, Norway
\item \Idef{org1145}Dipartimento di Fisica dell'Universit\`{a} and Sezione INFN, Cagliari, Italy
\item \Idef{org1270}Dipartimento di Fisica dell'Universit\`{a} and Sezione INFN, Padova, Italy
\item \Idef{org1315}Dipartimento di Fisica dell'Universit\`{a} and Sezione INFN, Trieste, Italy
\item \Idef{org1132}Dipartimento di Fisica dell'Universit\`{a} and Sezione INFN, Bologna, Italy
\item \Idef{org1285}Dipartimento di Fisica dell'Universit\`{a} `La Sapienza' and Sezione INFN, Rome, Italy
\item \Idef{org1154}Dipartimento di Fisica e Astronomia dell'Universit\`{a} and Sezione INFN, Catania, Italy
\item \Idef{org1290}Dipartimento di Fisica `E.R.~Caianiello' dell'Universit\`{a} and Gruppo Collegato INFN, Salerno, Italy
\item \Idef{org1312}Dipartimento di Fisica Sperimentale dell'Universit\`{a} and Sezione INFN, Turin, Italy
\item \Idef{org1103}Dipartimento di Scienze e Innovazione Tecnologica dell'Universit\`{a} del Piemonte Orientale and Gruppo Collegato INFN, Alessandria, Italy
\item \Idef{org1114}Dipartimento Interateneo di Fisica `M.~Merlin' and Sezione INFN, Bari, Italy
\item \Idef{org1237}Division of Experimental High Energy Physics, University of Lund, Lund, Sweden
\item \Idef{org1192}European Organization for Nuclear Research (CERN), Geneva, Switzerland
\item \Idef{org1227}Fachhochschule K\"{o}ln, K\"{o}ln, Germany
\item \Idef{org1122}Faculty of Engineering, Bergen University College, Bergen, Norway
\item \Idef{org1136}Faculty of Mathematics, Physics and Informatics, Comenius University, Bratislava, Slovakia
\item \Idef{org1274}Faculty of Nuclear Sciences and Physical Engineering, Czech Technical University in Prague, Prague, Czech Republic
\item \Idef{org1229}Faculty of Science, P.J.~\v{S}af\'{a}rik University, Ko\v{s}ice, Slovakia
\item \Idef{org1184}Frankfurt Institute for Advanced Studies, Johann Wolfgang Goethe-Universit\"{a}t Frankfurt, Frankfurt, Germany
\item \Idef{org1215}Gangneung-Wonju National University, Gangneung, South Korea
\item \Idef{org1212}Helsinki Institute of Physics (HIP) and University of Jyv\"{a}skyl\"{a}, Jyv\"{a}skyl\"{a}, Finland
\item \Idef{org1203}Hiroshima University, Hiroshima, Japan
\item \Idef{org1329}Hua-Zhong Normal University, Wuhan, China
\item \Idef{org1254}Indian Institute of Technology, Mumbai, India
\item \Idef{org36378}Indian Institute of Technology Indore (IIT), Indore, India
\item \Idef{org1266}Institut de Physique Nucl\'{e}aire d'Orsay (IPNO), Universit\'{e} Paris-Sud, CNRS-IN2P3, Orsay, France
\item \Idef{org1277}Institute for High Energy Physics, Protvino, Russia
\item \Idef{org1249}Institute for Nuclear Research, Academy of Sciences, Moscow, Russia
\item \Idef{org1320}Nikhef, National Institute for Subatomic Physics and Institute for Subatomic Physics of Utrecht University, Utrecht, Netherlands
\item \Idef{org1250}Institute for Theoretical and Experimental Physics, Moscow, Russia
\item \Idef{org1230}Institute of Experimental Physics, Slovak Academy of Sciences, Ko\v{s}ice, Slovakia
\item \Idef{org1127}Institute of Physics, Bhubaneswar, India
\item \Idef{org1275}Institute of Physics, Academy of Sciences of the Czech Republic, Prague, Czech Republic
\item \Idef{org1139}Institute of Space Sciences (ISS), Bucharest, Romania
\item \Idef{org27399}Institut f\"{u}r Informatik, Johann Wolfgang Goethe-Universit\"{a}t Frankfurt, Frankfurt, Germany
\item \Idef{org1185}Institut f\"{u}r Kernphysik, Johann Wolfgang Goethe-Universit\"{a}t Frankfurt, Frankfurt, Germany
\item \Idef{org1177}Institut f\"{u}r Kernphysik, Technische Universit\"{a}t Darmstadt, Darmstadt, Germany
\item \Idef{org1256}Institut f\"{u}r Kernphysik, Westf\"{a}lische Wilhelms-Universit\"{a}t M\"{u}nster, M\"{u}nster, Germany
\item \Idef{org1246}Instituto de Ciencias Nucleares, Universidad Nacional Aut\'{o}noma de M\'{e}xico, Mexico City, Mexico
\item \Idef{org1247}Instituto de F\'{\i}sica, Universidad Nacional Aut\'{o}noma de M\'{e}xico, Mexico City, Mexico
\item \Idef{org23333}Institut of Theoretical Physics, University of Wroclaw
\item \Idef{org1308}Institut Pluridisciplinaire Hubert Curien (IPHC), Universit\'{e} de Strasbourg, CNRS-IN2P3, Strasbourg, France
\item \Idef{org1182}Joint Institute for Nuclear Research (JINR), Dubna, Russia
\item \Idef{org1143}KFKI Research Institute for Particle and Nuclear Physics, Hungarian Academy of Sciences, Budapest, Hungary
\item \Idef{org1199}Kirchhoff-Institut f\"{u}r Physik, Ruprecht-Karls-Universit\"{a}t Heidelberg, Heidelberg, Germany
\item \Idef{org20954}Korea Institute of Science and Technology Information, Daejeon, South Korea
\item \Idef{org1160}Laboratoire de Physique Corpusculaire (LPC), Clermont Universit\'{e}, Universit\'{e} Blaise Pascal, CNRS--IN2P3, Clermont-Ferrand, France
\item \Idef{org1194}Laboratoire de Physique Subatomique et de Cosmologie (LPSC), Universit\'{e} Joseph Fourier, CNRS-IN2P3, Institut Polytechnique de Grenoble, Grenoble, France
\item \Idef{org1187}Laboratori Nazionali di Frascati, INFN, Frascati, Italy
\item \Idef{org1232}Laboratori Nazionali di Legnaro, INFN, Legnaro, Italy
\item \Idef{org1125}Lawrence Berkeley National Laboratory, Berkeley, California, United States
\item \Idef{org1234}Lawrence Livermore National Laboratory, Livermore, California, United States
\item \Idef{org1251}Moscow Engineering Physics Institute, Moscow, Russia
\item \Idef{org1140}National Institute for Physics and Nuclear Engineering, Bucharest, Romania
\item \Idef{org1165}Niels Bohr Institute, University of Copenhagen, Copenhagen, Denmark
\item \Idef{org1109}Nikhef, National Institute for Subatomic Physics, Amsterdam, Netherlands
\item \Idef{org1283}Nuclear Physics Institute, Academy of Sciences of the Czech Republic, \v{R}e\v{z} u Prahy, Czech Republic
\item \Idef{org1264}Oak Ridge National Laboratory, Oak Ridge, Tennessee, United States
\item \Idef{org1189}Petersburg Nuclear Physics Institute, Gatchina, Russia
\item \Idef{org1170}Physics Department, Creighton University, Omaha, Nebraska, United States
\item \Idef{org1157}Physics Department, Panjab University, Chandigarh, India
\item \Idef{org1112}Physics Department, University of Athens, Athens, Greece
\item \Idef{org1152}Physics Department, University of Cape Town, iThemba LABS, Cape Town, South Africa
\item \Idef{org1209}Physics Department, University of Jammu, Jammu, India
\item \Idef{org1207}Physics Department, University of Rajasthan, Jaipur, India
\item \Idef{org1200}Physikalisches Institut, Ruprecht-Karls-Universit\"{a}t Heidelberg, Heidelberg, Germany
\item \Idef{org1325}Purdue University, West Lafayette, Indiana, United States
\item \Idef{org1281}Pusan National University, Pusan, South Korea
\item \Idef{org1176}Research Division and ExtreMe Matter Institute EMMI, GSI Helmholtzzentrum f\"ur Schwerionenforschung, Darmstadt, Germany
\item \Idef{org1334}Rudjer Bo\v{s}kovi\'{c} Institute, Zagreb, Croatia
\item \Idef{org1298}Russian Federal Nuclear Center (VNIIEF), Sarov, Russia
\item \Idef{org1252}Russian Research Centre Kurchatov Institute, Moscow, Russia
\item \Idef{org1224}Saha Institute of Nuclear Physics, Kolkata, India
\item \Idef{org1130}School of Physics and Astronomy, University of Birmingham, Birmingham, United Kingdom
\item \Idef{org1338}Secci\'{o}n F\'{\i}sica, Departamento de Ciencias, Pontificia Universidad Cat\'{o}lica del Per\'{u}, Lima, Peru
\item \Idef{org1316}Sezione INFN, Trieste, Italy
\item \Idef{org1271}Sezione INFN, Padova, Italy
\item \Idef{org1313}Sezione INFN, Turin, Italy
\item \Idef{org1286}Sezione INFN, Rome, Italy
\item \Idef{org1146}Sezione INFN, Cagliari, Italy
\item \Idef{org1133}Sezione INFN, Bologna, Italy
\item \Idef{org1115}Sezione INFN, Bari, Italy
\item \Idef{org1155}Sezione INFN, Catania, Italy
\item \Idef{org1322}Soltan Institute for Nuclear Studies, Warsaw, Poland
\item \Idef{org36377}Nuclear Physics Group, STFC Daresbury Laboratory, Daresbury, United Kingdom
\item \Idef{org1258}SUBATECH, Ecole des Mines de Nantes, Universit\'{e} de Nantes, CNRS-IN2P3, Nantes, France
\item \Idef{org1304}Technical University of Split FESB, Split, Croatia
\item \Idef{org1168}The Henryk Niewodniczanski Institute of Nuclear Physics, Polish Academy of Sciences, Cracow, Poland
\item \Idef{org17361}The University of Texas at Austin, Physics Department, Austin, TX, United States
\item \Idef{org1173}Universidad Aut\'{o}noma de Sinaloa, Culiac\'{a}n, Mexico
\item \Idef{org1296}Universidade de S\~{a}o Paulo (USP), S\~{a}o Paulo, Brazil
\item \Idef{org1149}Universidade Estadual de Campinas (UNICAMP), Campinas, Brazil
\item \Idef{org1239}Universit\'{e} de Lyon, Universit\'{e} Lyon 1, CNRS/IN2P3, IPN-Lyon, Villeurbanne, France
\item \Idef{org1205}University of Houston, Houston, Texas, United States
\item \Idef{org20371}University of Technology and Austrian Academy of Sciences, Vienna, Austria
\item \Idef{org1222}University of Tennessee, Knoxville, Tennessee, United States
\item \Idef{org1310}University of Tokyo, Tokyo, Japan
\item \Idef{org1318}University of Tsukuba, Tsukuba, Japan
\item \Idef{org21360}Eberhard Karls Universit\"{a}t T\"{u}bingen, T\"{u}bingen, Germany
\item \Idef{org1225}Variable Energy Cyclotron Centre, Kolkata, India
\item \Idef{org1306}V.~Fock Institute for Physics, St. Petersburg State University, St. Petersburg, Russia
\item \Idef{org1323}Warsaw University of Technology, Warsaw, Poland
\item \Idef{org1179}Wayne State University, Detroit, Michigan, United States
\item \Idef{org1260}Yale University, New Haven, Connecticut, United States
\item \Idef{org1332}Yerevan Physics Institute, Yerevan, Armenia
\item \Idef{org15649}Yildiz Technical University, Istanbul, Turkey
\item \Idef{org1301}Yonsei University, Seoul, South Korea
\item \Idef{org1327}Zentrum f\"{u}r Technologietransfer und Telekommunikation (ZTT), Fachhochschule Worms, Worms, Germany
\end{Authlist}
\end{flushleft}
\endgroup

\end{appendix}
\end{document}